\documentclass[apj,twocolumn]{emulateapj}

\newcommand{\simgt}%
{\,\hbox{\lower0.6ex\hbox{$\sim$}\llap{\raise0.6ex\hbox{$>$}}}\,}
\newcommand{\simlt}%
{\,\hbox{\lower0.6ex\hbox{$\sim$}\llap{\raise0.6ex\hbox{$<$}}}\,}
\shorttitle{Non-Linear Bias, BAOs and Millennium Simulation}
\shortauthors{JEONG $\&$ KOMATSU}

\begin{document}
\title{%
  Perturbation Theory Reloaded II:
  Non-linear Bias, Baryon Acoustic Oscillations and Millennium Simulation
  In Real Space 
}%
\author{Donghui Jeong and Eiichiro Komatsu}
\affil{Department of Astronomy, University of Texas at Austin, 
       1 University Station, C1400, Austin, TX, 78712, USA}
\email{djeong@astro.as.utexas.edu}

\begin{abstract}
We calculate the non-linear galaxy power spectrum in real space, including non-linear
 distortion of the Baryon
 Acoustic Oscillations, using the standard 3rd-order
 perturbation theory (PT). The calculation is based upon the assumption that
 the number density of galaxies is a local function of the underlying,
 non-linear density field. The galaxy bias is allowed to be both
 non-linear and stochastic. We show that the PT calculation agrees with 
 the galaxy power spectrum estimated from the Millennium Simulation, in
 the weakly non-linear regime (defined by the matter power spectrum) at
 high redshifts, $1\le z\le6$. We also 
 show that, once 3 free parameters characterizing galaxy bias are
 marginalized over, the PT power spectrum fit to the Millennium
 Simulation data yields unbiased estimates of the distance scale, 
 $D$, to within the statistical error. This distance scale 
 corresponds to the angular diameter distance, $D_A(z)$, and 
 the expansion rate, $H(z)$, in real galaxy surveys.
 Our results presented
 in this paper are still restricted to real space. The future work should
 include the effects of non-linear redshift space
 distortion. Nevertheless, our results indicate that non-linear galaxy
 bias in the weakly non-linear
 regime at high redshifts is reasonably under control. 
\end{abstract}
\section{Introduction}
Surveys of galaxies are the oldest way of mapping cosmological
fluctuations. Over the last three decades they have been used for
measuring cosmological parameters, 
such as the matter density of the universe, $\Omega_m$
\citep[see][for a review]{peebles:POPC}. 

The galaxy surveys are largely complementary to CMB, as they allow us to
determine the important cosmological parameters that remain poorly
constrained by the CMB data alone
\citep[e.g.,][]{takada/komatsu/futamase:2006}: e.g., the mass of
neutrinos, 
the shape of the primordial power spectrum, and the properties of
dark energy. 

The latest data sets, Two Degree Field Galaxy Redshift
Survey \citep[2dFGRS,][]{cole/etal:2005} and  Sloan Digital Sky
Survey \citep[SDSS,][]{tegmark/etal:2006}, have enabled us to determine
   most of the cosmological parameters to
 better than 5\% accuracy,
when combined with the Cosmic Microwave Background (CMB) data from the Wilkinson
Microwave Anisotropy Probe \citep{bennett/etal:2003,spergel/etal:2003,spergel/etal:2007,hinshaw/etal:prep,dunkley/etal:prep,komatsu/etal:prep}.

The galaxy power spectrum, the Fourier transform of the galaxy two point
correlation function, has been used widely for extracting cosmological
information from the galaxy survey data.
The amplitude, overall shape, as well as oscillatory features (called
the Baryon Acoustic Oscillations, or BAOs) contain a wealth of
cosmological information \citep[see][for a recent review]{weinberg:COS}.
In order to extract this information correctly,
we must understand how the observed galaxy power spectra are related to
the underlying cosmological models. 

How do we model the galaxy power spectrum?
We may use the cosmological perturbation theory (PT). 
The accuracy of the linear PT has been verified
observationally by the
temperature and polarization data of CMB measured by WMAP 
\citep{hinshaw/etal:2003,hinshaw/etal:2007,kogut/etal:2003,page/etal:2007,nolta/etal:prep}. 
However, we cannot use the linear PT for 
the galaxy power spectrum, as the matter density field grows non-linearly
due to gravitational instability. One must therefore use the {\it
non-linear} PT.

There are three sources of non-linearities:
\begin{itemize}
\item[(1)] Non-linear evolution of the underlying matter density
field, which alters the matter power spectrum away from the linear prediction. 
\item[(2)] Non-linear galaxy bias, or non-linear mapping between the underlying matter density field 
and the distribution of collapsed objects such as dark matter halos and
galaxies, which alters the galaxy power spectrum
away from the matter power spectrum. 
\item[(3)] Non-linear redshift space distortion, which arises as
the observed redshifts of galaxies used for measuring
locations of galaxies along the line of sight contain both the Hubble
expansion and the peculiar velocity of galaxies. This leads to the
systematic shifts in the line-of-sight positions of galaxies, altering 
the galaxy power spectrum in redshift space away from that in real
space. 
\end{itemize}

Using the 3rd-order PT
\citep[see][for a review]{bernardeau/etal:2002} we have shown that
the first effect can be modeled accurately in the weakly non-linear
regime \citep[][hereafter Paper I]{jeong/komatsu:2006}.
In this paper we address the second effect, the non-linear
galaxy bias, using the 3rd-order PT. We will address
the third effect, the non-linear redshift space distortion, in the
future work. 

Our study is motivated by recently proposed high redshift galaxy 
surveys such as Cosmic Inflation Probe (CIP)\footnote{\sf
http://cfa-www.harvard.edu/cip}, 
Hobby-Eberly Dark Energy Experiment \citep[HETDEX;][]{hill/etal:2004},
Baryon Oscillation Spectroscopic Survey (BOSS)\footnote{\sf http://howdy.physics.nyu.edu/index.php/BOSS},
and Wide-field Fiber-fed Multi Object Spectrograph survey
\citep[WFMOS;][]{glazebrook/etal:prep}, to mention a few.
These proposed surveys will observe the galaxy power spectra to the
unprecedented precision, which demands the precision modeling of the
galaxy power spectrum at 1\% accuracy or better. 

Over the last decade, the non-linear PT, including modeling of 
non-linear galaxy power spectra, had been studied actively
\citep[see][for a review]{bernardeau/etal:2002}. 
However, PT had never been applied to the real data such as 2dFGRS or
SDSS, as non-linearities are too strong for PT to be valid at low
redshifts, $z<1$ \citep[e.g.,][]{meiksin/white/peacock:1999}.
At high redshifts, i.e., $z>1$, however, PT is expected to perform better
because of weaker non-linearity. 
In Paper I we have shown that the matter power spectrum
computed from the 3rd-order PT describes that from $N$-body simulations
accurately.\footnote{See also \citet{jain/bertschinger:1994} for
the earlier, pioneering work.} 

But, what about the {\it galaxy} power spectrum?
One may generally expect that, since non-linearities were milder in a 
high-$z$ universe, there should be a plenty of room for PT to be a good
approximation. On the other hand, galaxies were more highly biased at
higher redshifts for a given mass, and therefore one might suspect,
somewhat naively, that
non-linear bias could compromise the success of PT. 
In this paper we shall show that is not the case, and PT does provide a good
approximation to the galaxy power spectrum at high redshifts. 

This paper is organized as follows.
In \S~\ref{sec:PT} we give the formula for the 3rd-order PT galaxy power
spectrum. 
In \S~\ref{sec:DM} we compare the 3rd-order PT matter power spectrum
with the matter power spectrum estimated from the Millennium
Simulation \citep{springel/etal:2005}, in order to confirm our previous
results (Paper I) with the Millennium Simulation.
In \S~\ref{sec:galaxy} we show that the PT calculation of the galaxy
power spectrum agrees with the galaxy power spectrum estimated from the
Millennium Simulation in the weakly non-linear regime (defined by the
matter power spectrum) at high redshifts, $1\le z\le 6$.
In \S~\ref{sec:bias} we extract the distance scale from
the Millennium Simulation, which is related to the angular diameter distance
and the expansion rate of the universe in real surveys.
In \S~\ref{sec:conclusion} we give discussion and conclusions.

\section{Non-linear galaxy power spectrum from perturbation theory}
\label{sec:PT}
\subsection{Locality Assumption}
Galaxies are biased tracers of the underlying density field 
\citep{kaiser:1984}, which implies that the distribution of galaxies
depends on the underlying matter density fluctuations in a complex
way. This relation must depend
upon the detailed galaxy formation processes, which are not yet
understood completely. 

However, on large enough scales, one may approximate  this function as
a local function of the underlying density fluctuations, i.e., 
 the number density of galaxies at a given position in the universe is
 given solely by the underlying matter density at the same position.
With this approximation, one may expand the density fluctuations of
galaxies, $\delta_g$, in terms of the underlying matter density
fluctuations, as \citep{fry/gaztanaga:1993,mcdonald:2006}
\begin{equation}\label{eq:Taylor_expansion}
\delta_g(\mathbf{x})
=
\epsilon+
b_1\delta(\mathbf{x}) + \frac12b_2\delta^2(\mathbf{x})+\frac16b_3\delta^3(\mathbf{x})+\dots,
\end{equation}
where $b_n$ are the galaxy bias parameters, and 
$\epsilon$ is a random variable that represents the ``stochasticity'' of
the galaxy bias, i.e., the relation between $\delta_g(\mathbf{x})$ and 
$\delta(\mathbf{x})$ is not deterministic, but contains some noise
\citep[e.g.,][and references therein]{yoshikawa/etal:2001}.
We assume that the stochasticity is white noise, and is uncorrelated
with the density fluctuations, i.e., $\langle \epsilon\delta
\rangle=0$. While both of these assumptions should be violated at some small
scales, we assume that these are valid assumptions on the scales that we
are interested in -- namely, on the scales where the 3rd-order PT
describes the non-linear matter power spectrum with 1\% accuracy.
Since both bias parameters and stochasticity evolve in time
\citep{fry:1996,tegmark/peebles:1998}, we allow them to depend on redshifts.

One obtains the traditional ``linear bias model'' when the Taylor series
expansion 
given in Eq.~(\ref{eq:Taylor_expansion}) is truncated at the first
order and the stochasticity is ignored. 

The precise values of the galaxy bias parameters depend on the galaxy
formation processes, and different types of galaxies have different
galaxy bias parameters. 
However, we are {\it not} interested in the precise values of the galaxy bias
parameters, but only interested in extracting
cosmological parameters from the observed galaxy power spectra with {\it
all the bias parameters marginalized over}. 

\subsection{3rd-order PT galaxy power spectrum}
The analysis in this paper adopts the framework of \cite{mcdonald:2006},
and we briefly summarize the result for clarity.
We shall use the 3rd-order PT; thus, 
we shall keep the terms up to the 3rd order in $\delta$. 
The resulting power spectrum can be written in terms of
the linear matter power spectrum, $P_L(k)$, and the 3rd order matter
power spectrum, $P_{\delta\delta}(k)$, as
\begin{equation}\label{eq:3rd_PT_Pk}
P_{g}(k)
=
P_0 + \tilde{b}_1^2
\biggl[
P_{\delta\delta}(k) + \tilde{b}_2 P_{b2}(k) + \tilde{b}_2^2 P_{b22}(k)
\biggl],
\end{equation}
where $P_{b2}$ and $P_{b22}$ are given by 
\begin{displaymath}\label{eq:P_b2}
P_{b2}=
2
\int \frac{d^3 \mathbf{q}}{(2\pi)^3}
P_L(q) P_L(|\mathbf{k}-\mathbf{q}|)
F_2^{(s)}(\mathbf{q},\mathbf{k}-\mathbf{q}),
\end{displaymath}
and
\begin{displaymath}\label{eq:P_b22}
P_{b22}=
\frac{1}{2}\int \frac{d^3 \mathbf{q}}{(2\pi)^3}
P_L(q)
\biggl[
P_L(|\mathbf{k}-\mathbf{q}|) -P(q)
\biggl],
\end{displaymath}
respectively, with $F_2^{(2)}$ given by
\begin{displaymath}
F_2^{(s)}(\mathbf{q}_1,\mathbf{q}_2)
=
\frac{5}{7}+\frac{2}{7}
\frac{(\mathbf{q}_1\cdot\mathbf{q}_2)^2}{q_1^2q_2^2}
+
\frac{\mathbf{q}_2\cdot\mathbf{q}_2}{2}
\left(
	\frac{1}{q_1^2}+\frac{1}{q_2^2}
\right).
\end{displaymath}

We use the standard formula for $P_{\delta\delta}$ 
(see Eq.~(14) of Paper I and references therein).
Here, $\tilde{b}_1$, $\tilde{b}_2$, and $P_0$ are the 
non-linear bias parameters\footnote{These parameters correspond to
$b_1$, $b_2$, and $N$ in the original paper
by \citet{mcdonald:2006}.}, which are given in terms of the original
coefficients for the Taylor expansion as
\begin{eqnarray}
\label{eq:tb1b2}
\tilde{b}_1^2
&=&b_1^2+b_1b_3\sigma^2+\frac{68}{21}b_1b_2\sigma^2,\nonumber\\
\tilde{b}_2
&=&\frac{b_2}{\tilde{b}_1},\\
P_0 
&=& \langle \epsilon^2\rangle + \frac12b_2^2\int
 \frac{k^2dk}{2\pi^2}P_L^2(k)\nonumber, 
\end{eqnarray}
where $\sigma$ is the r.m.s. of density fluctuations.

We will never have to deal with the original coefficients, $b_1$, $b_2$,
$b_3$, or $\epsilon$.\footnote{For the expression of $P_g(k)$ with the
original coefficients, see \citet{heavens/matarrese/verde:1998,smith/scoccimarro/sheth:2007}.}  
Instead, we will only use the re-parametrized bias
parameters, $\tilde{b}_1$, $\tilde{b}_2$, and $P_0$, as these are
related more directly to the observables.
As shown by  \citet{mcdonald:2006},
in the large-scale limit, $k\rightarrow 0$, one finds 
\begin{equation}
 P_{g}(k)
\rightarrow 
P_0 + \tilde{b}_1^2 P_L(k).
\end{equation}
Therefore, in the large-scale limit one recovers the traditional linear
bias model plus the constant term.
Note that $\tilde{b}_1$ is the same as what is called the ``effective
bias'' in \cite{heavens/matarrese/verde:1998}. 

Throughout this paper we shall use Eq.~(\ref{eq:3rd_PT_Pk}) for calculating
the non-linear galaxy power spectra.

\subsection{Why we do not care about the precise values of bias parameters}
The precise values of the galaxy bias parameters depend on the details of the
galaxy formation and evolution, as well as on galaxy types,
luminosities, and so on. 

However, our goal is to extract the cosmological information from the
observed galaxy power spectra, without having to worry about which
galaxies we are using as tracers of the underlying density field.

Therefore, we will marginalize the likelihood function over the bias
parameters, without ever paying attention to their precise values. 
Is this approach sensible?

One might hope that one should be able to calculate the bias parameters 
for given properties of galaxies from the first principles using, e.g.,
sophisticated numerical simulations.

Less numerically expensive way of doing the same thing would be to use
the semi-analytical halo model approach, calibrated with a smaller set of
numerical simulations \citep[see][for a
review]{cooray/sheth:2002}. 
Using the peak-background split method \citep{sheth/tormen:1999} 
based upon the excursion set approach
\citep{Bond/etal:1991},  
one can calculate $b_1$, $b_2$, $b_3$, etc., the coefficients of the
Taylor series expansion given in Eq.~(\ref{eq:Taylor_expansion}), for the
density of dark matter halos. 
Once the bias parameters for dark matter halos are specified, 
the galaxy bias parameters may be calculated using the so-called Halo 
Occupation Distribution (HOD) \citep{seljak:2000}.

\citet{smith/scoccimarro/sheth:2007} have attempted this approach, and 
shown that it is  difficult to calculate even the power spectrum of
dark matter halos that matches $N$-body simulations.
The halo-model predictions for bias parameters are not yet accurate
enough, and we do not yet have a correct model for $P_0$. 

The situation would be even worse for the galaxy power spectrum, as
 we would have to model the HOD in addition to the halo bias.
At the moment the form of HOD is basically a free empirical function.
We therefore feel that it is  dangerous to rely on our limited understanding
 of these complications for computing the bias parameters. 

This is the reason why we have decided to give up predicting the precise
values of bias parameters entirely.  
Instead, we shall treat 3 bias parameters, $\tilde{b}_1$, $\tilde{b}_2$,
and $P_0$, as free parameters, and fit them to the observed galaxy power
spectra simultaneously with the cosmological parameters.

The most important question that we must ask is the following, ``using the
3rd-order PT 
with 3 bias parameters, can we extract the correct cosmological
parameters from the galaxy power spectra?'' If the answer is yes, we
will not have to worry about the precise values of bias parameters
anymore.

\section{Dark Matter Power spectrum from Millennium Simulation}
\label{sec:DM}
In this section we show that the matter power spectrum computed from the
3rd-order PT agrees with that estimated from the Millennium
Simulation \citep{springel/etal:2005}. This result confirms our previous
finding (Paper I).

Using the result obtained in this
section we define the maximum wavenumber, $k_{max}$, below which the
3rd-order PT may be trusted. The matter power spectrum gives an unambiguous
definition of $k_{max}$, which will then be used thereafter when we
analyze power spectra of halos and galaxies in \S~\ref{sec:galaxy}.

\subsection{Millennium Simulation}
The Millennium simulation \citep{springel/etal:2005} is a large $N$-body simulation with the box
size of $(500~\textrm{Mpc}/h)^3$ and $2160^3$ dark matter particles. The
cosmological parameters used  
in the simulation are 
$(\Omega_{dm},\Omega_{b},\Omega_\Lambda,h)
=(0.205,~0.045,~0.75,~0.73)$. 

The primordial power spectrum used in the simulation is 
the scale-invariant Peebles-Harrison-Zel'dovich spectrum, $n_s=1.0$, and
the linear r.m.s. density fluctuation smoothed with a top-hat filter of
radius $8~h^{-1}\mathrm{Mpc}$ is $\sigma_8=0.9$.
Note that these values are significantly larger than 
the latest values found from the WMAP 5-year data, 
$\sigma_8\simeq 0.8$ and $n_s\simeq 0.96$
\citep{dunkley/etal:prep,komatsu/etal:prep}, which implies that
non-linearities in the 
Millennium Simulation should be stronger than those in our Universe.

The Millennium Simulation was carried out using the {\sf GADGET} code
\citep{springel/yoshida/white:2001,springel:2005}.
The {\sf GADGET} uses the tree Particle Mesh (tree-PM) gravity solver,
which tends to have a larger dynamic range than the traditional PM
solver for the same box size
and the same number of particles (and meshes)\citep{heitmann/etal:prep}.
Therefore, the matter power spectrum from the Millennium Simulation does
not suffer from an artificial suppression of power as much as those from
the PM codes.

The initial particle distribution was generated at the initial redshift
of $z_{ini}=127$ using the standard Zel'dovich approximation. 
While the initial conditions generated from the standard Zel'dovich
approximation tend to produce an artificial suppression of power at
later times, and the higher-order scheme such as the second-order
Lagrangian perturbation theory usually produces better results
\citep{scoccimarro:1998,crocce/pueblas/scoccimarro:2006}, 
the initial redshift of the Millennium Simulation, $z_{ini}=127$, is
reasonably high for the resulting power spectra to have converged in the
weakly non-linear regime.

The mass of each dark matter particle in the simulation is
$M_{dm}=8.6\times 10^8 M_\odot/h$. They require at least 20 particles
per halo for their halo finder, and thus the minimum mass resolution of
halos is given by 
$M_{halo}\ge20M_{dm}\simeq 1.7 \times 10^{10}~M_\odot/h$.
Therefore, the Millennium Simulation covers the mass range that is
relevant to real galaxy surveys that would detect galaxies with masses
in the range of $M\simeq 10^{11}-10^{12}~M_\odot$. 
This property distinguishes our study from the previous studies on 
non-linear distortion of BAOs due to galaxy bias
\citep[e.g.,][]{smith/scoccimarro/sheth:2007,huff/etal:2007}, whose
mass resolution was greater than $\sim 10^{12}~M_\odot$.

In addition to the dark matter halos, the Millennium database\footnote{
\textsf{http://www.g-vo.org/MyMillennium2/}} 
also provides galaxy catalogues from two different 
semi-analytic galaxy formation models
\citep{delucia/blaizot:2007,croton/etal:2006,bower/etal:2006,benson/etal:2003,cole/etal:2000}.
These catalogues give us an excellent opportunity for testing 
validity  of the non-linear
galaxy power spectrum model based upon the 3rd-order PT
with the unprecedented precision. 

\subsection{3rd-order PT versus Millennium Simulation: Dark Matter Power
  Spectrum}
First, we compare the matter power spectrum from the Millennium
simulation with the 3rd-order PT calculation. The matter power
spectrum we use here was  measured directly from the  Millennium 
simulation on the fly.\footnote{We thank Volker Springel for providing
us with the matter power spectrum data.}

\begin{figure}
\centering
\rotatebox{90}{%
  \includegraphics[width=6.5cm]{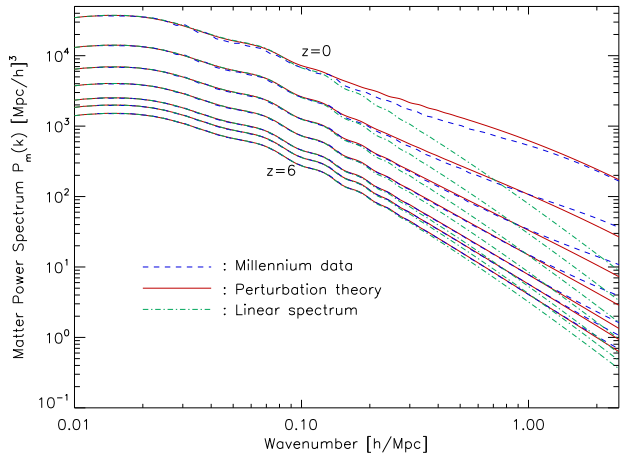}
}%
\caption{%
Matter power spectrum at $z=0$, 1, 2, 3, 4, 5 and 6 
(\textit{from top to bottom}) derived from the Millennium 
Simulation (\textit{dashed lines}),
the 3rd-order PT (\textit{solid lines}), and the linear PT
(\textit{dot-dashed lines}). 
}%
\label{fig1}
\end{figure}
\begin{figure}
\centering
\rotatebox{90}{%
  \includegraphics[width=6.5cm]{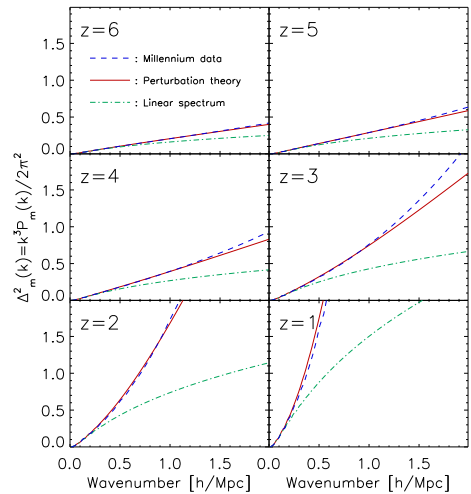}
}%
\caption{%
Dimensionless matter power spectrum, $\Delta^2(k)$,
at $z=1$, 2, 3, 4, 5, and 6.
The dashed and solid lines show the Millennium Simulation data and the
 3rd-order PT calculation, respectively.
The dot-dashed lines show the linear power spectrum.
}%
\label{fig2}
\end{figure}
\begin{figure}
\centering
\rotatebox{90}{
	\includegraphics[width=6.5cm]{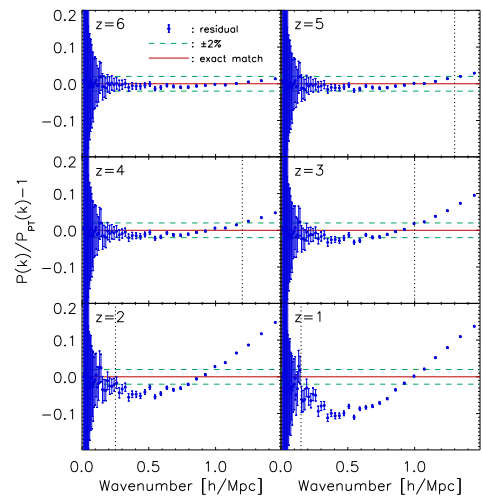}
}
\caption{
Fractional difference between the matter power spectra from the 3rd-order PT
 and that from the Millennium Simulation, $P_m^{sim}(k)/P_m^{PT}-1$
(dots with errorbars). The solid lines show the perfect match, 
while the  dashed lines show $\pm 2\%$ accuracy.
We also show $k_{max}(z)$, below which we trust the prediction from the 
3rd-order PT, as a vertical dotted line.
}%
\label{fig3}
\end{figure}

\begin{deluxetable}{ccc}


\tabletypesize{\footnotesize}


\tablecaption{Maximum wavenumbers, $k_{max}$, for the Millennium Simulation}
\tablenum{1}
\label{table:kmax}

\tablehead{\colhead{$z$} & \colhead{$k_{max}$} & \colhead{$\tilde{k}_{max}$} \\ 
\colhead{} & \colhead{($h/\mathrm{Mpc}$)} & \colhead{$(h/\mathrm{Mpc})$} } 

\startdata
6 & 1.5 & 1.99 \\
5 & 1.3 & 1.37 \\
4 & 1.2 & 1.02 \\
3 & 1.0 & 0.60 \\
2 & 0.25 & 0.35 \\
1 & 0.15 & 0.20 
\enddata


\tablecomments{$z$: redshift\\
$k_{max}$: the maximum wavenumber for the simulated $P_m(k)$ to 
agree with the PT calculation at 2\% accuracy within the statistical
 error of the Millennium Simulation\\
$\tilde{k}_{max}$: 
$\tilde{k}_{max}$ is defined by $\Delta^2_m(\tilde{k}_{max})=0.4$
which is the criteria recommended in Paper I.}

\end{deluxetable}
\begin{figure}
\centering
\rotatebox{90}{
	\includegraphics[width=6.5cm]{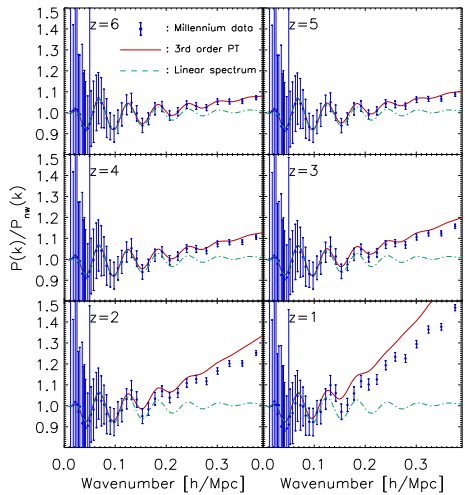}
}
\caption{
Distortion of BAOs due to non-linear matter clustering. All of the power
 spectra have been divided by 
a smooth power spectrum without baryonic oscillations from eq.
(29) of \citet{eisenstein/hu:1998}. The error bars show the simulation
 data, while the solid lines show the PT calculations. The dot-dashed 
lines show the linear theory calculations.
The power spectrum data shown here have been taken from Figure~6 of
 \citet{springel/etal:2005}.
}%
\label{fig4}
\end{figure}

Figure \ref{fig1} shows the matter power spectrum from the Millennium 
simulation (dashed lines), the 3rd-order PT calculation 
(solid lines), and the linear PT (dot-dashed lines)
for seven different redshifts, $z=0$, 1, 2, 3, 4, 5, and 6.
The analytical calculation of the 3rd-order
PT reproduces the non-linear matter power spectrum from the 
Millennium Simulation accurately at high redshifts, i.e., $z>1$,
up to certain maximum wavenumbers, $k_{max}$, that will be specified below.
To facilitate the comparison better, we show the dimensionless 
matter power spectrum, $\Delta_m^2(k)\equiv k^3P_m(k)/2\pi^2$, 
in Figure \ref{fig2}. 

We find the maximum wavenumber, $k_{max}(z)$, below which we trust 
the prediction from the 3rd-order PT, by comparing the matter 
power spectrum from PT and the Millennium Simulation. 
The values of $k_{max}$ found here will be used later 
when we analyze the halo/galaxy power spectra.

In Paper I we have defined $k_{max}$ such that the fractional difference
between PT and the average of $\sim 100$ simulations is 1\%. 
Here, we have only one realization, and thus the results are subject to 
statistical fluctuations that might be peculiar to this particular
realization. 
Therefore, we relax our criteria for $k_{max}$:
we define $k_{max}$ such that the fractional difference
between PT and the Millennium Simulation is 2\%.

Figure \ref{fig3} shows the fractional differences 
at $z=1$, 2, 3, 4, 5, and 6. Since we have only one realization, 
we cannot compute  statistical errors from the standard deviation of
multiple realizations.  
Therefore, we derive errors from the leading-order 4-point function
assuming Gaussianity of the underlying density fluctuations (see
Appendix~\ref{sec:appA}), 
$\sigma_{P(k)}=P(k)/\sqrt{N_k}$,
where $N_k$ is the  number of independent Fourier modes per bin at a
given $k$ shown in Figure~\ref{fig3}.

We give the values of $k_{max}$ in Table \ref{table:kmax}. We shall use
these values when we fit the halo/galaxy power spectrum in the
next section. Note that $k_{max}$ decreases rapidly below $z=2$.
It is because $P(k)/P_{PT}(k)-1$ is not a monotonic function of $k$.
The dip in $P(k)/P_{PT}(k)-1$ is larger than $2\%$ at lower redshift, 
$z<2$, while it is inside of the $2\%$ range at $z\ge3$.
Therefore, our criteria of $2\%$ make that sudden change.
This feature is due to the limitation of the standard 3rd order PT. 
However, we can remove this feature by using the improved perturbation theory,
e.g. using renormalization group techniques. 
(See, Figure 9 of \citet{matarrese/pietroni:2007}.)

We also give the values of $\tilde{k}_{max}$, for which
$\Delta^2_m(\tilde{k}_{max})=0.4$ (criteria recommended in Paper I).
The difference between $k_{max}$ and $\tilde{k}_{max}$ is probably due
to the fact that we have only one realization of the Millennium
Simulation, and thus estimation of $k_{max}$ is noisier.
Note that the values of $\tilde{k}_{max}$ given in Table
 \ref{table:kmax} 
are smaller than those given in Paper I. 
This is simply because $\sigma_8$ of the Millennium Simulation
 ($\sigma_8=0.9$) is larger than that of Paper I ($\sigma_8=0.8$).

In Figure \ref{fig4} we show 
the matter power spectra divided a smooth spectra without BAOs
\citep[Eq.~(29) of][]{eisenstein/hu:1998}. 
The results are consistent with what we have found in Paper I:
although BAOs in the matter power spectrum are distorted heavily
by non-linear evolution of matter fluctuations, 
the analytical predictions from the 3rd-order PT
capture the distortions very well at high redshifts, $z>2$. 

At lower redshifts, $z\sim 1$, the 3rd-order PT is clearly
insufficient, and one needs to go beyond the standard PT. 
This is a subject of recent studies
\citep{crocce/scoccimarro:2008,matarrese/pietroni:2007,taruya/hiramatsu:2008,valageas:2007,matsubara:2008,mcdonald:2007}.

\section{HALO/GALAXY POWER SPECTRUM AND THE NON-LINEAR BIAS MODEL}
\label{sec:galaxy}
In this section we compare the 3rd-order PT galaxy power spectrum 
with the power spectra of dark matter halos and galaxies
estimated from the Millennium Simulation.
After briefly describing the analysis method in \S~\ref{sec:analysis},
we  analyze the halo bias and galaxy bias in \S~\ref{sec:halo}
and \S~\ref{sec:gal}, respectively. We then study 
the dependence of bias parameters on halo/galaxy mass in
\S~\ref{sec:mass_dependence}. 

\subsection{Analysis method}
\label{sec:analysis}
\begin{deluxetable*}{cccccccc}


\tabletypesize{\footnotesize}


\tablecaption{Summary of six snapshots from the Millennium Simulation}

\tablenum{2}
\label{table:Msummary}
\tablehead{\colhead{$z$} & \colhead{$z_\mathrm{show}$} & \colhead{$N_h$} & \colhead{$1/n_h$} & \colhead{$N_{Mg}$} & \colhead{$1/n_{Mg}$} & \colhead{$N_{Dg}$} & \colhead{$1/n_{Dg}$} \\ 
 \colhead{} &\colhead{} & \colhead{} & \colhead{($[\mathrm{Mpc}/h]^3$)} & \colhead{} & \colhead{($[\mathrm{Mpc}/h]^3$)} & \colhead{} & \colhead{($[\mathrm{Mpc}/h]^3$)} } 

\startdata
5.724&6 & 5,741,720 & 21.770 & 6,267,471 & 19.944 & 4,562,368 & 27.398 \\
4.888&5 & 8,599,981 & 14.535 & 9,724,669 & 12.854 & 7,604,063 & 16.439 \\
4.179&4 & 11,338,698 & 11.024 & 13,272,933 & 9.418 & 10,960,404 & 11.405 \\
3.060&3 & 15,449,221 & 8.091 & 19,325,842 & 6.468 & 17,238,935 & 7.251 \\
2.070&2 & 17,930,143 & 6.972 & 23,885,840 & 5.233 & 22,962,129 & 5.444 \\
1.078&1 & 18,580,497 & 6.727 & 26,359,329 & 4.742 & 27,615,058 & 4.527
\enddata


\tablecomments{$z$: the exact redshift of each snapshot\\ 
$z_\mathrm{show}$: the redshift we quote in this paper\\
$N_h$: the number of MPA halos in each snapshot; $1/n_h$: the
 corresponding Poisson shot noise\\
$N_{Mg}$: the number of MPA galaxies in each snapshot; $1/n_{Mg}$: the
 corresponding Poisson shot noise\\
$N_{Dg}$: the number of Durham galaxies in each snapshot; $1/n_{Dg}$: the
 corresponding Poisson shot noise}


\end{deluxetable*}

We choose six redshifts between $1\le z\le6$ from 63 snapshots of 
the Millennium Simulation, and use all the available catalog of halos 
(MPA Halo (MHalo), hereafter `halo') and two galaxy catalogues
(MPA Galaxies, hereafter `Mgalaxy'; Durham Galaxies,
hereafter `Dgalaxy') at each redshift. The 
exact values of redshifts and the other relevant information
of chosen snapshots are summarized in Table \ref{table:Msummary}.

Halos are the groups of matter particles found directly from 
the Millennium Simulation. First, 
the dark matter groups (called FOF group) 
are identified by using Friends-of-Friends (FoF) algorithm with 
a linking length equal to 0.2 of the mean particle separation.
Then, each FoF group is divided into the gravitationally 
bound local overdense regions, which we call halos here.

Mgalaxies and Dgalaxies are the galaxies assigned to the halos
using two different semi-analytic galaxy formation codes:
L-Galaxies \citep[Mgalaxies,][]{delucia/blaizot:2007,croton/etal:2006} 
and GALFORM \citep[Dgalaxies,][]{bower/etal:2006,benson/etal:2003,cole/etal:2000}. 

While both models successfully explain a number of observational 
properties of galaxies like the break shape of 
the galaxy luminosity function, star formation rate, etc, they differ
in detailed implementation.
For example, while the L-Galaxies code uses the halo merger tree constructed 
by MHalos, the GALFORM code uses different 
criteria for identifying subhalos inside the FOF group, 
and thus uses a different merger tree.
Also, two models use different gas cooling prescriptions
and different initial mass functions (IMF) of star formation:
L-Galaxies and GALFORM define the cooling radius, 
within which gas has a sufficient time to cool,
by comparing the cooling time with halo dynamical time and the 
age of the halo, respectively.
Cold gas turns into stars with two different IMFs:
the L-Galaxies code ueses IMF from \citet{chabrier:2003} and 
the GALFORM code uses \citet{kennicutt:1983}.
In addition to that, they treat AGN (Active Galactic Nucleus) 
feedback differently: the L-Galaxies code 
introduces a parametric model of AGN feedback depending 
on the black hole mass and the virial velocity of halo, 
and the GALFORM code imposes the condition
that cooling flow is quenched when the energy released by radiative cooling
(cooling luminosity) is less than some fraction 
(which is modeled by a parameter, $\epsilon_{\mathrm{SMBH}}$) 
of Eddington luminosity of the black hole.
For more detailed comparison of the two model, we refer readers to the
original papers cited above.

We compute the halo/galaxy power spectra from the Millennium Simulation
as follows:
\begin{itemize}
 \item [(1)] Use the Cloud-In-Cell (CIC) 
mass distribution scheme to 
calculate the density field on $1024^3$ 
regular grid points from each catalog. 
 \item [(2)] Fourier-transform the discretized density field using {\sf
       FFTW}\footnote{{\sf http://www.fftw.org}}.
 \item [(3)] Deconvolve the effect of the CIC pixelization and aliasing
       effect. We divide $P(\mathbf{k},z)\equiv|\delta(\mathbf{k},z)|^2$ 
at each cell by the following
window function \citep{Jing:2005}:
\begin{equation}\label{eq:window}
W(\mathbf{k})=\prod_{i=1}^3 
\left[
1-\frac{2}{3}\sin^2 \left(\frac{\pi k_i}{2 k_{N}}\right)
\right],
\end{equation}
where  $\mathbf{k}=(k_1,k_2,k_3)$, and $k_N\equiv\pi/H$
is the Nyquist frequency, 
($H$ is the physical size of the grid).\footnote{Note that
       Eq.~(\ref{eq:window})  is strictly valid 
for the flat (white noise) power spectrum, $P(k)={\rm constant}$.
Nevertheless, it is still accurate for our purposes because, on small
scales, both the halo and galaxy power spectra are dominated by 
the shot noise, which is also given by $P(k)={\rm constant}$.}
 \item [(4)]  Compute $P(k,z)$ by taking the angular average of
CIC-corrected $P(\mathbf{k},z)\equiv|\delta(\mathbf{k},z)|^2$ within a 
spherical shell defined by
$k-\Delta k/2 < |\mathbf{k}| < k+\Delta k/2$.
Here, $\Delta k=2\pi/500~[h/\rm{Mpc}]$ is the
fundamental frequency that corresponds to the box size of the Millennium
Simulation. 
\end{itemize}

From the measured power spectra we find the maximum likelihood
values of the bias parameters using the likelihood 
function approximated as a Gaussian:
\begin{equation}\label{eq:likelihood}
\mathcal{L}(\tilde{b}_1,\tilde{b}_2,P_0)
=\prod_{k_i<k_{max}}
\frac{1}{\sqrt{2\pi\sigma_{Pi}^2}}
\exp
\left[-
\frac{
(P_{obs,i}-P_{g,i})^2
}{2\sigma_{Pi}^2}
\right], 
\end{equation}
where $k_i$'s are integer multiples of the fundamental frequency
$\Delta k$, $P_{obs,i}$ is the measured power spectrum at $k=k_i$,
$P_{g,i}$ is the theoretical model given by Eq.~(\ref{eq:3rd_PT_Pk}),
and $\sigma_{Pi}$ is the statistical error in the measured power
spectrum.

We estimate  $\sigma_{Pi}$ in the same way as in \S~\ref{sec:DM}
(see also Appendix~\ref{sec:appA}). However, the power spectrum of 
the point-like particles like halos and galaxies includes the Poisson shot
noise, $1/n$, where $n$ is the number density of objects, on top of the
power spectrum due to clustering.
Therefore, $\sigma_{Pi}$ must also include the shot-noise
contribution. We use
\begin{equation}\label{eq:varpk}
\sigma_{Pi}=\sigma_P(k_i)=\sqrt{\frac{1}{N_{ki}}}\left[P_g(k_i)+\frac{1}{n}\right],
\end{equation}
where 
\begin{equation}
N_{ki}=2\pi\left(\frac{k}{\Delta k}\right)^2
\end{equation} 
is the number of independent Fourier modes used for estimating the 
power spectrum and $P_g(k_i)$ is the halo/galaxy power spectrum at $k=k_i$.
Here, $\Delta k=2\pi/(500~h^{-1}~\rm{Mpc})$ is the fundamental wavenumber
of the Millennium Simulation.
Note that we subtract the Poisson shot noise contribution,
$P_{shot}=1/n$, from the observed power spectrum 
before the likelihood analysis.

Eq.~(\ref{eq:varpk}) shows that the error on $P_{obs}(k)$ depends upon
the underlying $P_g(k)$. For the actual data analysis one should vary
$P_g(k)$
in the numerator of Eq.~(\ref{eq:likelihood}) as well as that in
$\sigma_{Pi}$, simultaneously. However, to simplify the analysis, we evaluate the
likelihood function in an iterative way: we first find the best-fitting
$P_g(k)$ using $\sigma_{Pi}$ with $P_g(k)$ in Eq.~(\ref{eq:varpk})
replaced by $P_{obs}(k)$. Let us call this $\tilde{P}_g(k)$. We then use 
$\tilde{P}_g(k)$ in  Eq.~(\ref{eq:varpk}) for finding the best-fitting 
$P_g(k)$ that we shall report in this paper. 
Note that we iterate this procedure only once for current study.

Finally, we compute the 1-d marginalized 1-$\sigma$ interval (or the
marginalized $68.27\%$ confidence
interval) of each bias parameter by integrating the likelihood function
(Eq.~(\ref{eq:likelihood})), assuming   a  flat prior on the 
bias parameters (see also Appendix \ref{sec:appB}). 

We first analyze the power spectrum of halos (in \S~\ref{sec:halo}) as
well as that of galaxies (in \S~\ref{sec:gal})
using all the halos and all the galaxies in the Millennium halo/galaxy
catalogues. We then study the mass dependence of bias parameters
in \S~\ref{sec:mass_dependence}.

In order to show that the non-linear bias model (Eq.~\ref{eq:3rd_PT_Pk})
provides a much better fit than the linear bias model,
we also fit the measured power spectra with two linear bias models:
(i) linear bias with the linear matter power spectrum, and 
(ii) linear bias with the non-linear matter power spectrum from the
3rd-order PT. When fitting with the linear model, we use 
$k_{max}=0.15~[h/\mathrm{Mpc}]$ for all redshift bins.

\subsection{Halo power spectra}
\label{sec:halo}
\subsubsection{Measuring non-linear halo bias parameters}
\begin{figure*}
\centering
\rotatebox{90}{
	\includegraphics[width=10cm]{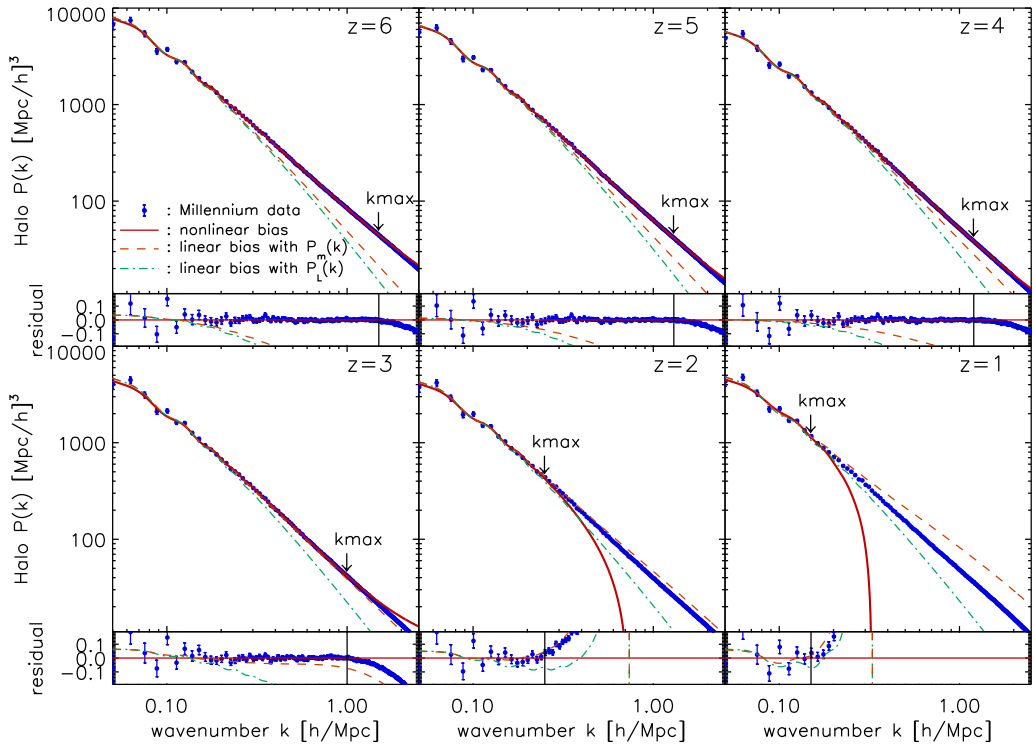}
}
\caption{
Halo power spectra from the Millennium Simulation at
$z=1$, 2, 3, 4, 5, and 6. Also shown in smaller panels 
are the residual of fits. The points with errorbars show
the measured halo power spectra, while the solid, dashed, and
dot-dashed lines
show the best-fitting non-linear bias model (Eq.~(\ref{eq:3rd_PT_Pk})),
the best-fitting linear bias with the non-linear matter power spectrum,
and the best-fitting linear bias with the linear matter power spectrum,
respectively. Both linear models have been fit for 
$k_{max,linear}=0.15~[h~\mathrm{Mpc}^{-1}]$, whereas 
$k_{max}(z)$ given in Table~\ref{table:kmax} (also marked in each panel)
 have been used  for the non-linear bias model.  
}%
\label{fig5}
\end{figure*}
\begin{deluxetable*}{ccclcccc}
%
\tabletypesize{\footnotesize}
%
\tablecaption{Non-linear halo bias parameters and the corresponding 68\%
 interval estimated from the MPA halo power spectra
}
%
\tablenum{3}
\label{table:halobias}
%
\tablehead{\colhead{$z$} & \colhead{$\tilde{b}_1$} & \colhead{$\tilde{b}_2$} & \colhead{$P_0$} & \colhead{$b_1^L$} & \colhead{$b_1^{LL}$} & \colhead{$b_1^{ST}$} & \colhead{$\tilde{b}_2^{ST}$} \\ 
\colhead{} & \colhead{} & \colhead{} & \colhead{($[\mathrm{Mpc}/h]^3$)} & \colhead{} & \colhead{} & \colhead{} & \colhead{} } 
%
\startdata
6 & 3.41$\pm$0.01 & 1.52$\pm$0.03 & 141.86$\pm$3.73 & 3.50$\pm$0.03 & 3.51$\pm$0.03 & 3.69 & 2.10 \\
5 & 2.76$\pm$0.01 & 0.91$\pm$0.03 & 57.77$\pm$2.84 & 2.79$\pm$0.03 & 2.80$\pm$0.03 & 3.16 & 1.70 \\
4 & 2.27$\pm$0.01 & 0.52$\pm$0.03 & 22.65$\pm$1.88 & 2.28$\pm$0.02 & 2.29$\pm$0.02 & 2.77 & 1.40 \\
3 & 1.52$\pm$0.01 & -1.94$\pm$0.05 & 329.42$\pm$10.6 & 1.62$\pm$0.01 & 1.63$\pm$0.01 & 2.23 & 1.07 \\
2 & 1.10$\pm$0.06 & -2.12$\pm$0.65 & 507.25$\pm$214.7 & 1.19$\pm$0.01 & 1.20$\pm$0.01 & 1.84 & 0.76 \\
1 & 0.74$\pm$0.09 & -3.05$\pm$1.49 & 1511.46$\pm$526.7 & 0.88$\pm$0.01 & 0.90$\pm$0.01 & 1.54 & 0.58 

\enddata
%
%
\tablecomments{$z$: redshift\\
$\tilde{b}_1$, $\tilde{b}_2$, $P_0$: non-linear bias parameters\\
$b_1^{L}$: linear bias parameter for the linear bias model with the 3rd-order
 matter power spectrum\\ 
$b_1^{LL}$: linear bias parameter for the linear bias model with the
 linear power spectrum\\ 
$b_1^{ST}$, $\tilde{b}_2^{ST}$: non-linear bias parameters calculated
 from the Sheth-Tormen model, $\tilde{b}_2^{ST}$=$b_2^{ST}/\tilde{b}_1$\\
\textit{
Caution: We estimate 1-$\sigma$ ranges for the low redshift ($z\le3$)
only for the peak which involves the maximum likelihood value. 
If two peaks in maginalized likelihood function are blended, we use
only unblended side of the peak to estimate the 1-$\sigma$ range.
}
}
%
%
\end{deluxetable*}
\begin{figure}
\centering
\rotatebox{90}{
	\includegraphics[width=7cm]{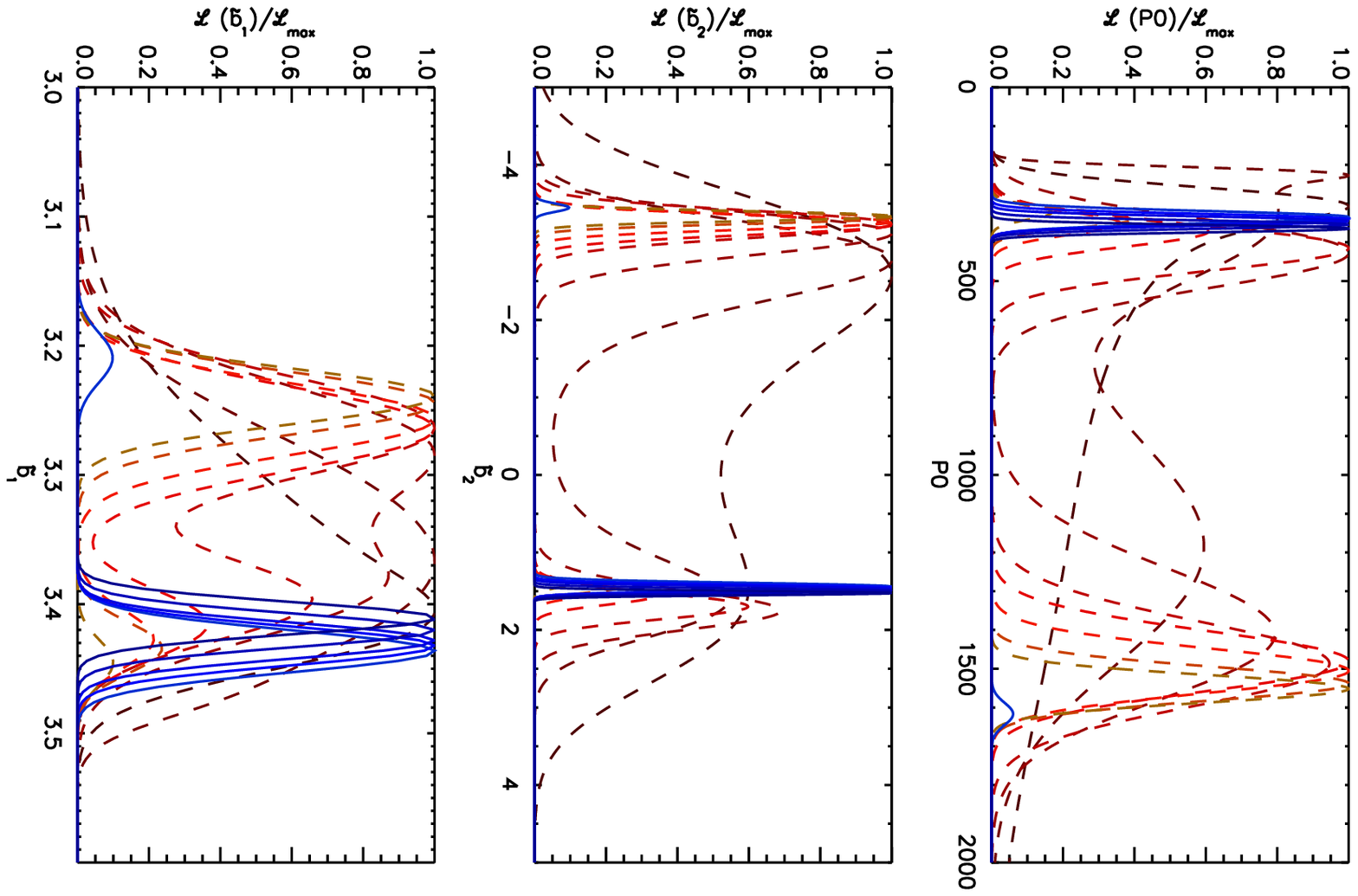}
}
\caption{
One-dimensional marginalized distribution of non-linear bias parameters
at $z=6$: from top to bottom panels, $P_0$, $\tilde{b}_2$, and $\tilde{b}_1$.
Different lines show the different values of $k_{max}$ used for the
 fits. The dashed and solid lines correspond to  
$0.3\le k_{max}/[h~\mathrm{Mpc}^{-1}]\le1.0$ and  $1.0<
 k_{max}/[h~\mathrm{Mpc}^{-1}]\le1.5$, respectively. The double-peak
 structure disappears for higher $k_{max}$.
}%
\label{fig6}
\end{figure}
\begin{figure}
\centering
\rotatebox{90}{
	\includegraphics[width=7cm]{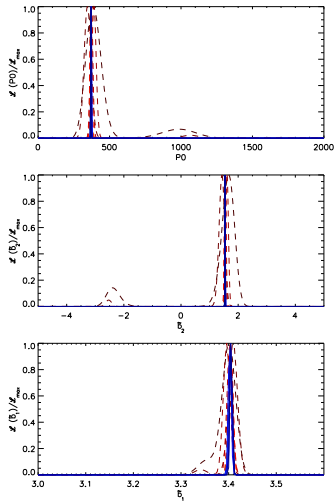}
}
\caption{
Same as Figure~\ref{fig6}, but for a Monte Carlo simulation
of a galaxy survey with  a bigger box size,  $L_{box}=1.5~\mathrm{Gpc}/h$. 
}%
\label{fig7}
\end{figure}
\begin{figure}
\centering
\rotatebox{90}{
	\includegraphics[width=7cm]{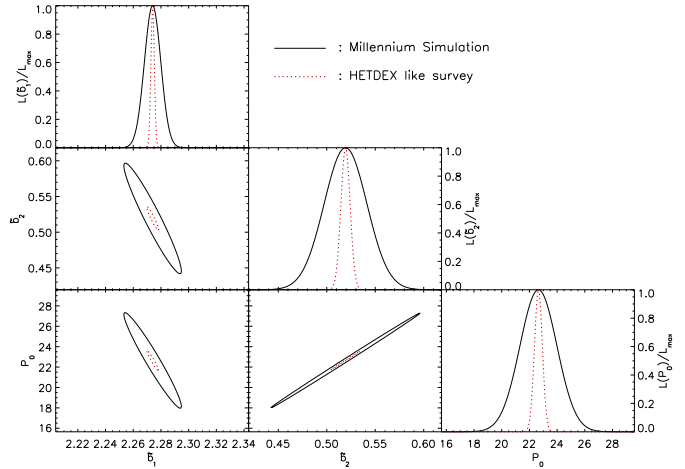}
}
\caption{
One-dimensional marginalized constraints and two-dimensional 
joint marginalized constraint of 2-$\sigma$ ($95.45\%$ CL) 
range for bias parameters
($\tilde{b}_1$,$\tilde{b}_2$,$P_0$). 
Covariance matrices are calculated from 
the Fisher information matrix 
(Eq. (\ref{eq:fisher_bias})) with the best-fitting bias 
parameters for halo at $z=4$.
}%
\label{fig8}
\end{figure}

\begin{figure*}
\centering
\rotatebox{90}{
	\includegraphics[width=10cm]{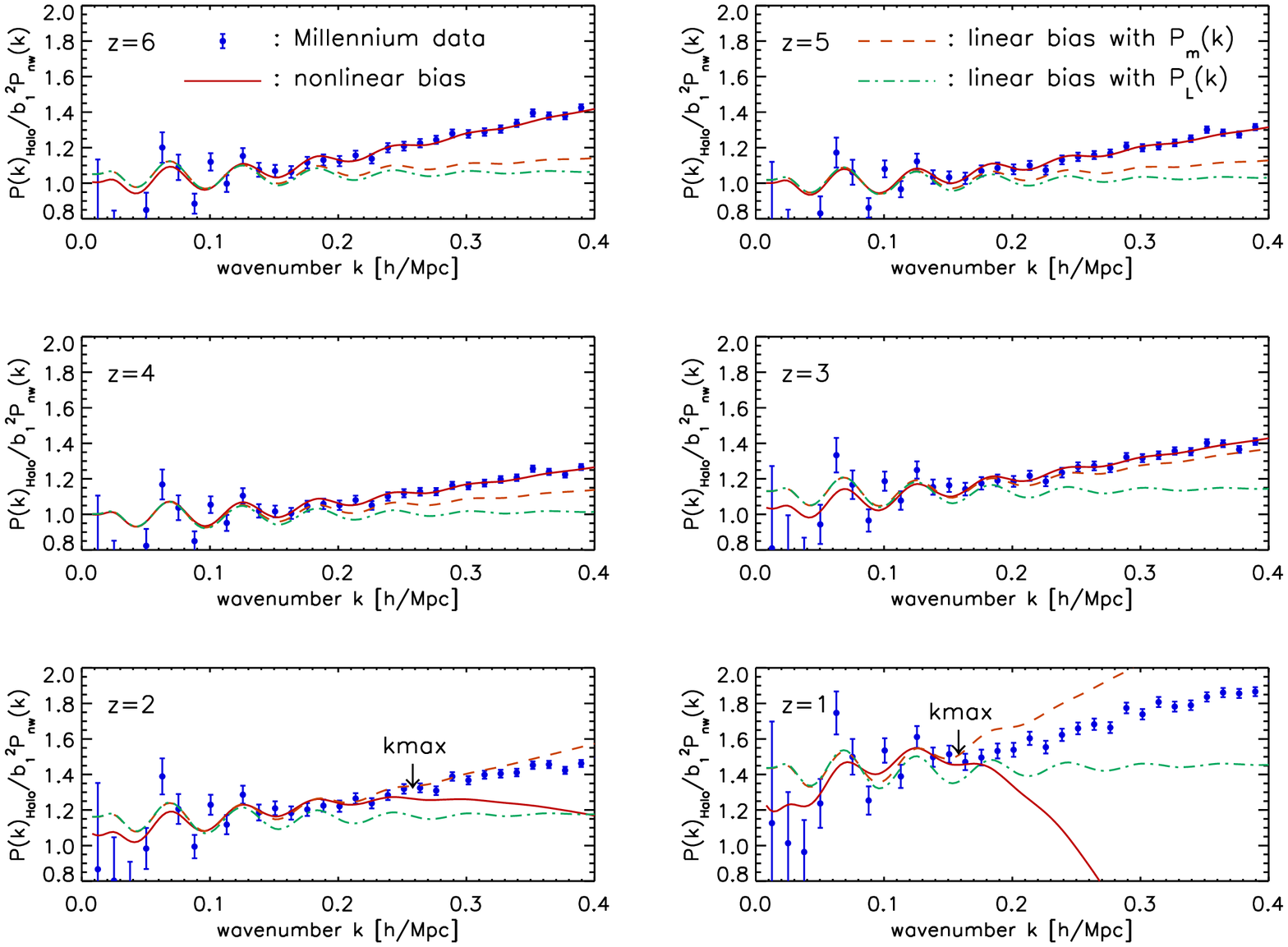}
}
\caption{
Distortion of BAOs due to non-linear matter clustering and non-linear halo bias.
All of the power spectra have been divided by a smooth power 
spectrum without baryonic oscillations from equation~(29) of 
\cite{eisenstein/hu:1998}. The errorbars show the Millennium Simulation, 
while the solid lines show the PT calculations.
The dashed lines show the linear bias model with the non-linear 
matter power spectrum, and the dot-dashed lines show the linear
bias model with the linear matter power spectrum.
Therefore, the difference between the solid lines and the dashed lines
 shows the distortion solely due to non-linear halo bias.
}%
\label{fig9}
\end{figure*}

Figure \ref{fig5} shows the best-fitting non-linear ({\it solid
lines}) and linear bias
models ({\it dashed and dot-dashed lines}), compared with the halo
spectra estimated from the Millennium Simulation ({\it points with
errorbars}). The smaller panels show the residuals of fits.
The maximum wavenumber used in the fits, $k_{max}(z)$, are also marked with
the arrows (bigger panels), and the vertical lines (smaller panels). 
We find that  the non-linear bias model provides substantially better 
fits than the linear bias models. 

We find that all of non-linear bias parameters, $\tilde{b}_1$,
$\tilde{b}_2$, and $P_0$, are strongly degenerate, when the maximum
wavenumbers used in the fits, $k_{max}$, are small. 
In Figure~\ref{fig6} we show the one-dimensional marginalized
distribution of bias parameters at $z=6$, as a function of 
$k_{max}$. For lower $k_{max}$, $0.3\le
k_{max}/[h~\mathrm{Mpc}^{-1}]\le1.0$, the marginalized distribution has two
peaks ({\it dashed lines}), indicating strong degeneracy with the other
parameters. The double-peak structure disappears for
$1.0< k_{max}/[h~\mathrm{Mpc}^{-1}]\le1.5$  ({\it solid lines}).

We find that the origin of degeneracy is simply due to the small box size of the
Millennium Simulation, i.e., the lack of statistics, or too a large
sampling variance. 
To show this,  
we have generated a mock Monte Carlo realization of 
halo power spectra, assuming a much bigger box size,
$L_{box}=1.5~h^{-1}~\mathrm{Gpc}$, which gives the fundamental frequency of
$\Delta k=5.0\times10^{-4}~h~\mathrm{Mpc}^{-1}$.
Note that this volume roughly corresponds to that would be surveyed by
the HETDEX survey \citep{hill/etal:2004}.
We have used the 
same non-linear matter power spectrum and the best-fitting bias
parameters from the Millennium Simulation (MPA halos) when creating Monte Carlo
realizations.  The resulting marginalized likelihood function at $z=6$
 is shown in Figure~\ref{fig7}.
The double-peak structure has disappeared even for
low $k_{max}$, $k_{max}= 0.3~h~\mathrm{Mpc}^{-1}$.
Therefore, we conclude that the double-peak problem
can be resolved simply by increasing the survey volume.

The best-fitting non-linear halo bias parameters and the corresponding
1-$\sigma$ intervals are summarized in Table \ref{table:halobias}.
Since we know that the double-peak structure is spurious, we pick one 
peak that corresponds to the maximum likelihood value, 
and quote the 1-$\sigma$ interval. 
At $z\le 2$, the bias parameters are not constrined very well
because of lower $k_{max}$ and the limited statistics of the 
Millennium Simulation, and hence the two peaks are blended; 
thus, we estimate 1-$\sigma$ range only from the unblended side of the 
marginalized likelihood function.
Two linear bias parameters, one with the linear matter power
spectrum and another with the non-linear PT matter power spectrum, are
also presented with their 1-$\sigma$ intervals. 

\subsubsection{Degeneracy of bias parameters}
In order to see how strongly degenerate bias parameters are,
we calculate the covariance matrix of each pair of bias 
parameters.
We calculate the covariance matrix of each pair of bias parameters
by using the Fisher information matrix, which is the inverse
of the covariance matrix.
The Fisher information matrix for the galaxy power spectrum 
can be approximated as
\citep{tegmark:1997c}
\footnote{Eq. (\ref{eq:fisher_bias}) is equivalent to Eq. (6)
in \citet{tegmark:1997c}. The number of $k$ mode in real space
power spectrum from a survey of volume $V$ is 
(See Appendix A for notations.)
$$
N_{k_n}=
\frac{4\pi k_n^2 \delta k_n}{2 (\delta k_n)^3}
=\frac{Vk_n^2\delta k_n}{4\pi^2}.
$$
Then, the variance of power spectrum (Eq. (\ref{eq:varpk})) becomes
$$
\sigma_P^2(k_n)
=
\frac{4\pi^2}{Vk_n^2\delta k_n}\left[P(k_n)+\frac{1}{n}\right]^2
=
\frac{4\pi^2P(k_n)^2}{k_n^2\delta k_n}\frac{1}{V_{\mathrm{eff}}(k_n)},
$$
where $V_{\mathrm{eff}}$
is the constant density version of Eq. (5) of \citet{tegmark:1997c}.
Finally, the elements of Fisher matrix are given by
\begin{eqnarray*}
F_{ij}
&=&\sum_n \frac{1}{\sigma_P^2(k_n)}
\frac{\partial P(k_n,\mathbf{\theta})}{\partial \theta_i}
\frac{\partial P(k_n,\mathbf{\theta})}{\partial \theta_j}\\
&=&
\frac{1}{4\pi^2}
\sum_n 
\frac{\partial P(k_n,\mathbf{\theta})}{\partial \theta_i}
\frac{\partial P(k_n,\mathbf{\theta})}{\partial \theta_j}
\frac{V_{\mathrm{eff}}(k_n)k_n^2\delta k_n}{P(k_n)^2}
\end{eqnarray*}
which is the same as Eq. (6) in \citet{tegmark:1997c}.
}
\begin{equation}
\label{eq:fisher_bias}
F_{ij}=\sum_n \frac{1}{\sigma_P^2(k_n)}
\frac{\partial P(k_n,\mathbf{\theta})}{\partial \theta_i}
\frac{\partial P(k_n,\mathbf{\theta})}{\partial \theta_j}
\end{equation}
where $\mathbf{\theta}$ is a vector in the parameter space,
$\theta_i=\tilde{b}_1$, $\tilde{b}_2$ ,$P_0$, for
$i=1$, $2$, $3$, respectively.
We calculate the marginalized errors on the bias parameters as following. 
We first calculate the full Fisher matrix and invert it to 
estimate the covariance matrix. Then, we get the 
the covariance matrices of any pairs of bias parameters by taking
the $2$ by $2$ submatrix of the full covariance matrix.
Figure \ref{fig8} shows the resulting 2-$\sigma$
($95.45\%$ interval) contour for the bias parameters at $z=4$.
We find the strong degeneracy between $\tilde{P}_0$
and $\tilde{b}_2$. We also find that 
$\tilde{b}_1$ is degenerate with the other two parameters.
On top of the error contours for the Millennium Simulation,
we show the expected contour from the HETDEX like survey
($1.5~\mathrm{Gpc}/h$). Since the volume of HETDEX like survey
is 27 times bigger, the likelihood functions and the error-contours
are about a factor of 5 smaller than those from the Millennium Simulation.
Other than that, two contours follow the same trend.
Results are the same for the other redshifts.

\subsubsection{Comparison with the halo model predictions}
The effective linear bias, $\tilde{b}_1$, is larger  at 
higher redshifts.
This is the expected result, as halos of mass greater than 
$\sim 10^{10} M_\odot$ were rarer in the earlier time,
resulting in the larger bias. 

From the same reason, we expect that 
the non-linear bias parameters, $\tilde{b}_2$  and $P_0$, are also
larger at higher $z$. 
While we observe the expected trend at $z\ge 4$, the results from 
$z\le 3$ are somewhat peculiar. 
This is probably due to the large sampling variance making 
the fits unstable: for $z\le 3$ the maximum wavenumbers inferred 
from the matter power spectra are less than $1.0~h~\mathrm{Mpc}^{-1}$
(see Table~\ref{table:kmax}), which makes the likelihood function
double-peaked and leaves the bias parameters poorly constrained. 
 
How do these bias parameters compare with the expected values?
We use the halo model for computing the mass-averaged bias parameters,
$b_1^{ST}$ and $b_2^{ST}$, assuming that the minimum mass 
is given by the minimum mass of the MPA halo catalogue,
$M_{min}=1.72\times10^{10} M_\odot/h$:
\begin{equation}
b_i^{ST}=\frac
{\int_{M_{min}}^{M_{max}}\frac{dn}{dM}M b_i(M) dM}
{\int_{M_{min}}^{M_{max}}\frac{dn}{dM}MdM},
\end{equation}
where $dn/dM$ is the Sheth-Tormen mass function and $b_i(M)$ is the
$i$-th order bias parameter from \citet{scoccimarro/etal:2001}.

There is one subtlety. The halo model predicts the coefficients
of the Taylor series (Eq.~(\ref{eq:Taylor_expansion})), whereas what we
have measured are the re-parametrized bias parameters given by
Eq.~(\ref{eq:tb1b2}).
However, the formula for $\tilde{b}_1$ includes the mass variance,
$\sigma^2$, which depends on our choice of a smoothing scale that is not
well defined. This shows how difficult it is to actually compute the
halo power spectrum from the halo model. While the measured values of
$\tilde{b}_1$ and the predicted $b_1^{ST}$ compare reasonably well, it
is clear that we cannot use the predicted bias values for doing cosmology.

For $\tilde{b}_2$, we compute $\tilde{b}_2^{ST}=b_2^{ST}/\tilde{b}_1$
where  $\tilde{b}_1$ is the best-fitting value from the Millennium
Simulation. This would give us a semi apple-to-apple comparison.
Nevertheless, while the agreement is reasonable at $z\ge 4$, the halo model
predictions should not be used for predicting $\tilde{b}_2$ either.

\subsubsection{Comments on the bispectrum}
\label{sec:bispectrum}
While the degeneracy between bias parameters may appear to be a serious
issue, there is actually a powerful way of breaking degeneracy:
the bispectrum, the Fourier transform of the 3-point correlation
function \citep{matarrese/verde/heavens:1997}. The reduced bispectrum, which is the bispectrum normalized
properly by the power spectrum, depends primarily on two bias
parameters, $\tilde{b}_1$ and $\tilde{b}_2$, nearly independent of the
cosmological parameters \citep{sefusatti/etal:2006}.
Therefore, one can use this property to fix the bias parameters,
and use the power spectrum for determining the cosmological parameters
and the remaining bias parameter, $P_0$. \citet{sefusatti/komatsu:2007}
have shown that the planned high-$z$ galaxy surveys would be able to
determine  $\tilde{b}_1$ and $\tilde{b}_2$ with a few percent accuracy.

We have begun studying the bispectrum of the Millennium Simulation.
Our preliminary results show that  we can indeed obtain
better constraints on $\tilde{b}_1$ and $\tilde{b}_2$ from the
bispectrum than from the power spectrum, provided that we use the same
$k_{max}$ for both the bispectrum and power spectrum analyses.
Therefore, even when the non-linear bias parameters are poorly 
constrained by the power spectrum alone, or have the double-peak
likelihood function from the power spectrum for lower $k_{max}$, we can
still find tight constraints on $\tilde{b}_1$ and $\tilde{b}_2$ from the
bispectrum. These results will be reported elsewhere.

\subsubsection{Effects on BAOs}
In Figure~\ref{fig9} we show the distortion of BAO features 
due to non-linear matter clustering and non-linear bias. 
To show only the distortions of BAOs at each redshift,
we have divided the halo power spectra by smooth power spectra without
baryonic oscillations from equation (29) of \citet{eisenstein/hu:1998}
with  $\tilde{b}_1^2$ multiplied.
Three theoretical models are shown: 
the non-linear bias model (\textit{solid line}),
a linear bias model with the 3rd-order matter power spectrum
(\textit{dashed line}), and a linear bias model with the linear matter
power spectrum  (\textit{dot-dashed line}).
Therefore, the difference between the solid lines and the dashed lines
is solely due to non-linear halo bias.

The importance of non-linear bias affecting BAOs grows with $z$;
however, as the matter clustering is weaker at higher $z$, the
 3rd-order PT still performs better than at lower $z$.
In other words, the higher bias at higher $z$ does not mean that surveys
at higher $z$ are worse at measuring BAOs; on the contrary, it is still
easier to model the halo power spectrum at higher $z$ than at lower
$z$. 
For $z\ge3$, where $k_{max}$ is larger than the BAO scale, 
the distortion of BAOs  is  modeled very well by 
the non-linear bias model, while  the linear bias models fail badly. 

The sampling variance of the Millennium Simulation at $k\lesssim
0.15~h~{\rm Mpc}^{-1}$ is too large for us to study the distortion on
the first two BAO peaks. Since the PT performs well at higher $k$,
we expect that the PT describes the first two peaks even better.
However, to show this explicitly one would need to run a bigger
simulation with a bigger volume with the same mass resolution as the
Millennium Simulation, which should be entirely doable with the existing
computing resources.

\subsection{Galaxy power spectra}
\label{sec:gal}
\subsubsection{Measuring non-linear galaxy bias parameters}
\begin{figure*}
\centering
\rotatebox{90}{
	\includegraphics[width=10cm]{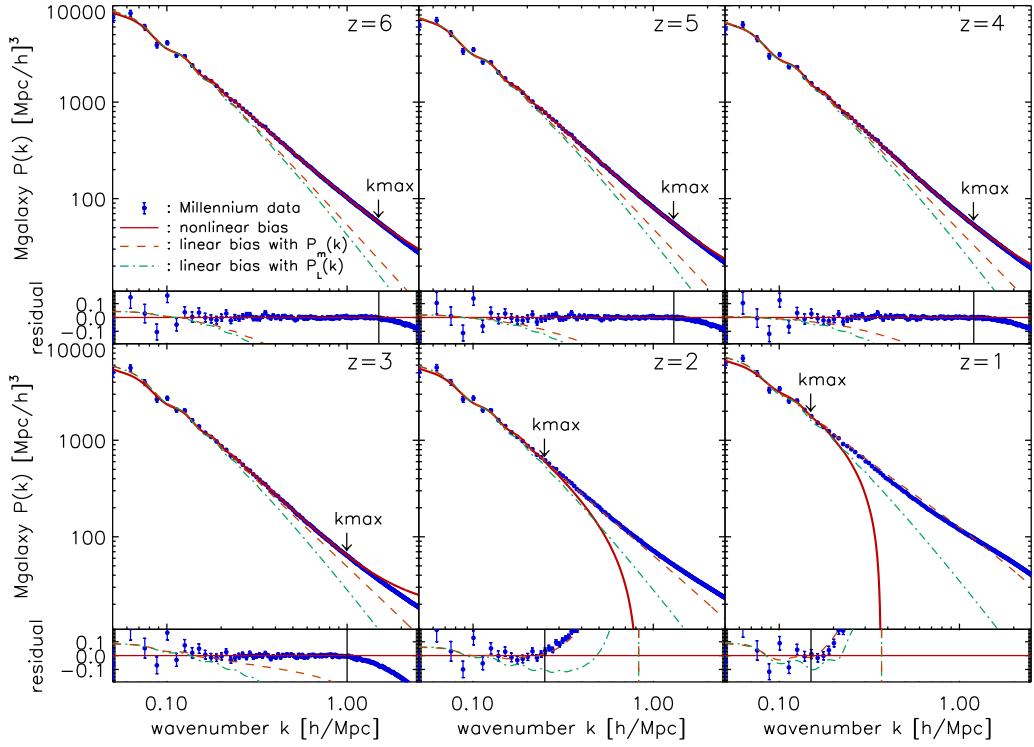}
}
\caption{
Same as Figure~\ref{fig5}, but for the MPA galaxy catalogue (Mgalaxy).
}%
\label{fig10}
\end{figure*}
\begin{figure*}
\centering
\rotatebox{90}{
	\includegraphics[width=10cm]{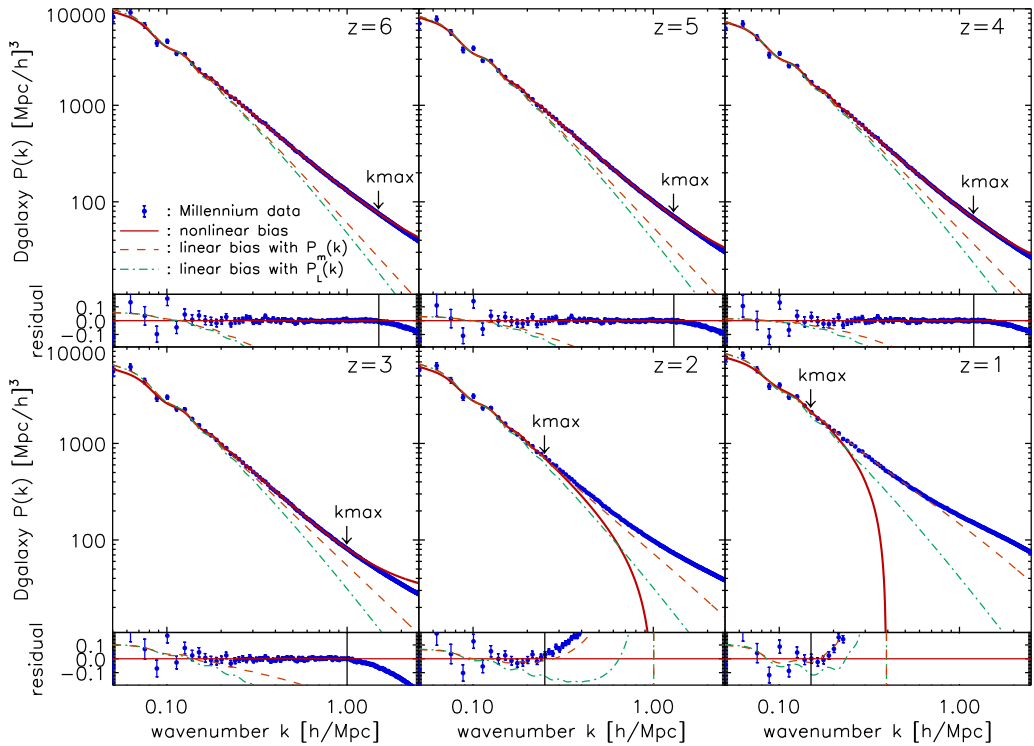}
}
\caption{
Same as Figure~\ref{fig5}, but for the Durham galaxy catalogue (Dgalaxy).
}%
\label{fig11}
\end{figure*}
\begin{deluxetable*}{ccclcccc}


\tabletypesize{\footnotesize}

\tablecaption{Non-linear halo bias parameters and the corresponding 68\%
 interval estimated from the MPA galaxy power spectra 
}

\tablenum{4}
\label{table:Mgbias}

\tablehead{\colhead{$z$} & \colhead{$\tilde{b}_1$} & \colhead{$\tilde{b}_2$} & \colhead{$P_0$} & \colhead{$b_1^L$} & \colhead{$b_1^{LL}$} & \colhead{$b_1^{ST}$} & \colhead{$\tilde{b}_2^{ST}$} \\ 
\colhead{} & \colhead{} & \colhead{} & \colhead{($[h/\mathrm{Mpc}]^3$)} & \colhead{} & \colhead{} & \colhead{} & \colhead{} } 

\startdata
6 & 3.55$\pm$0.01 & 1.70$\pm$0.03 & 194.23$\pm$4.45 & 3.67$\pm$0.03 & 3.68$\pm$0.03 & 3.10 & 1.03 \\
5 & 2.93$\pm$0.01 & 1.08$\pm$0.03 & 94.08$\pm$3.71 & 2.97$\pm$0.03 & 2.98$\pm$0.03 & 2.55 & 0.59 \\
4 & 2.46$\pm$0.01 & 0.68$\pm$0.03 & 47.79$\pm$2.84 & 2.47$\pm$0.02 & 2.48$\pm$0.02 & 2.13 & 0.28 \\
3 & 1.69$\pm$0.01 & -2.12$\pm$0.04 & 486.69$\pm$12.7 & 1.83$\pm$0.02 & 1.83$\pm$0.02 & 1.58 & -0.12 \\
2 & 1.28$\pm$0.08 & -2.16$\pm$0.64 & 738.22$\pm$291.3 & 1.40$\pm$0.01 & 1.40$\pm$0.01 & 1.19 & -0.34 \\
1 & 0.89$\pm$0.11 & -2.97$\pm$1.60 & 2248.35$\pm$786.13 & 1.09$\pm$0.01 & 1.10$\pm$0.01 & 0.91 & -0.45

\enddata


\tablecomments{$z$: redshift\\
$\tilde{b}_1$, $\tilde{b}_2$, $P_0$: non-linear bias parameters\\
$b_1^{L}$: linear bias parameter for the linear bias model with the 3rd-order
 matter power spectrum\\ 
$b_1^{LL}$: linear bias parameter for the linear bias model with the
 linear power spectrum\\ 
$b_1^{ST}$, $\tilde{b}_2^{ST}$: non-linear bias parameters calculated
 from the Sheth-Tormen model, $\tilde{b}_2^{ST}$=$b_2^{ST}/\tilde{b}_1$
\\
\textit{
Caution: We estimate 1-$\sigma$ ranges for the low redshift ($z\le3$)
only for the peak which involves the maximum likelihood value. 
If two peaks in maginalized likelihood function are blended, we use
only unblended side of the peak to estimate the 1-$\sigma$ range.
}
}

\end{deluxetable*}
\begin{deluxetable*}{ccclcccc}


\tabletypesize{\footnotesize}

\tablecaption{Non-linear halo bias parameters and the corresponding 68\%
 interval estimated from the Durham galaxy power spectra 
}

\tablenum{5}
\label{table:Dgbias}

\tablehead{\colhead{$z$} & \colhead{$\tilde{b}_1$} & \colhead{$\tilde{b}_2$} & \colhead{$P_0$} & \colhead{$b_1^L$} & \colhead{$b_1^{LL}$} & \colhead{$b_1^{ST}$} & \colhead{$\tilde{b}_2^{ST}$} \\ 
\colhead{} & \colhead{} & \colhead{} & \colhead{($[h/\mathrm{Mpc}]^3$)} & \colhead{} & \colhead{} & \colhead{} & \colhead{} } 

\startdata
6 & 3.73$\pm$0.01 & 1.96$\pm$0.03 & 288.39$\pm$5.82 & 3.90$\pm$0.04 & 3.90$\pm$0.04 & 3.10 & 0.98 \\
5 & 3.07$\pm$0.01 & 1.26$\pm$0.03 & 143.15$\pm$4.81 & 3.15$\pm$0.03 & 3.15$\pm$0.03 & 2.55 & 0.56 \\
4 & 2.57$\pm$0.01 & 0.83$\pm$0.03 & 78.97$\pm$3.93 & 2.60$\pm$0.02 & 2.61$\pm$0.02 & 2.13 & 0.26 \\
3 & 1.75$\pm$0.01 & -2.26$\pm$0.04 & 604.65$\pm$13.8 & 1.92$\pm$0.02 & 1.93$\pm$0.02 & 1.58 & -0.11 \\
2 & 1.36$\pm$0.08 & -2.14$\pm$0.65 & 843.49$\pm$331.4 & 1.49$\pm$0.01 & 1.50$\pm$0.01 & 1.19 & -0.32 \\
1 & 0.96$\pm$0.11 & -2.94$\pm$1.62 & 2640.20$\pm$960.32 & 1.18$\pm$0.01 & 1.20$\pm$0.01 & 0.91 & -0.42
\enddata


\tablecomments{$z$: redshift\\
$\tilde{b}_1$, $\tilde{b}_2$, $P_0$: non-linear bias parameters\\
$b_1^{L}$: linear bias parameter for the linear bias model with the 3rd-order
 matter power spectrum\\ 
$b_1^{LL}$: linear bias parameter for the linear bias model with the
 linear power spectrum\\ 
$b_1^{ST}$, $\tilde{b}_2^{ST}$: non-linear bias parameters calculated
 from the Sheth-Tormen model, $\tilde{b}_2^{ST}$=$b_2^{ST}/\tilde{b}_1$
\\
\textit{
Caution: We estimate 1-$\sigma$ ranges for the low redshift ($z\le3$)
only for the peak which involves the maximum likelihood value. 
If two peaks in maginalized likelihood function are blended, we use
only unblended side of the peak to estimate the 1-$\sigma$ range.
}
}

\end{deluxetable*}
\begin{figure*}
\centering
\rotatebox{90}{
	\includegraphics[width=10cm]{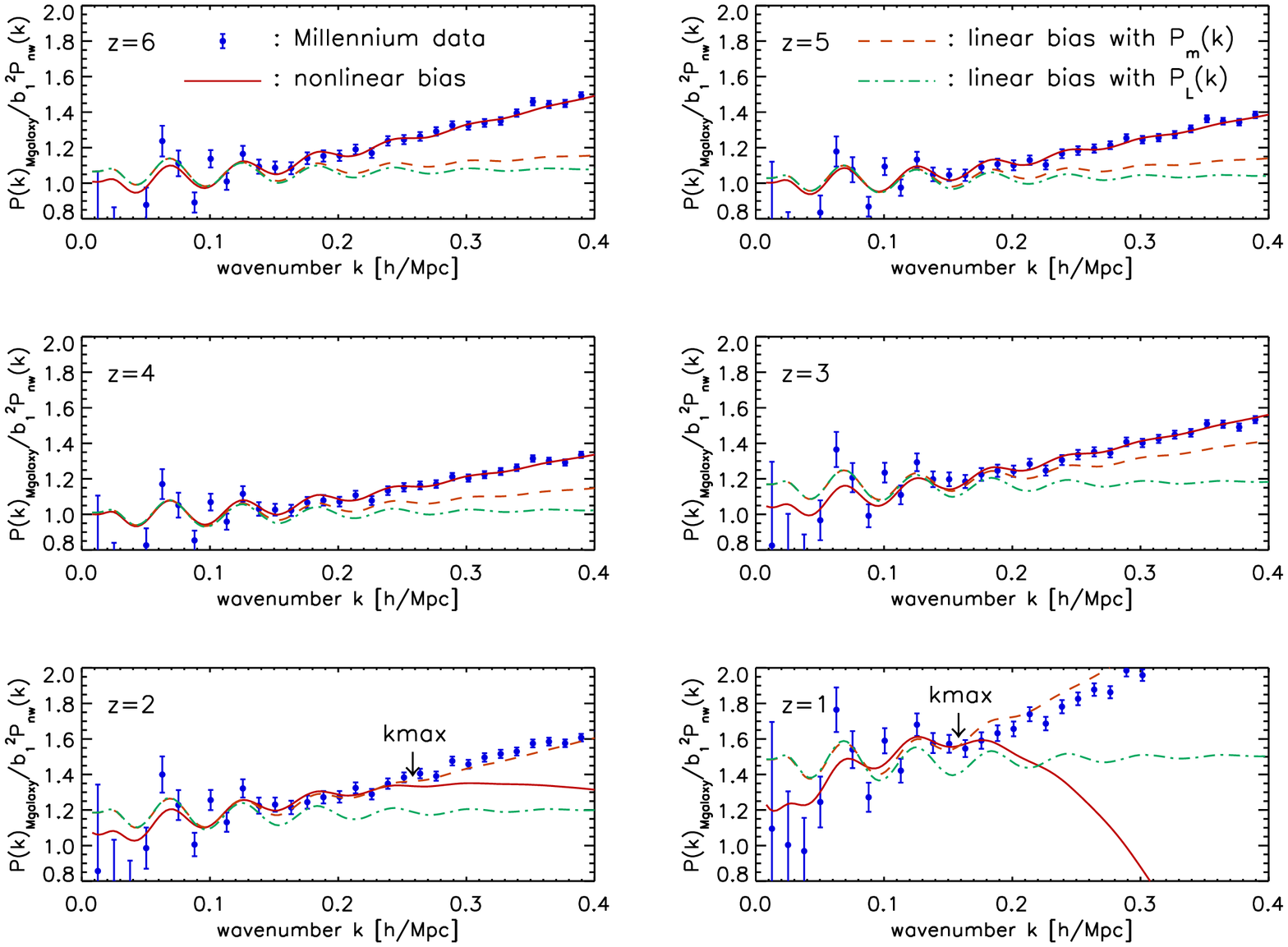}
}
\caption{
Same as Figure~\ref{fig9}, but for the MPA galaxy power spectrum
 (Mgalaxy). 
}%
\label{fig12}
\end{figure*}
\begin{figure*}
\centering
\rotatebox{90}{
	\includegraphics[width=10cm]{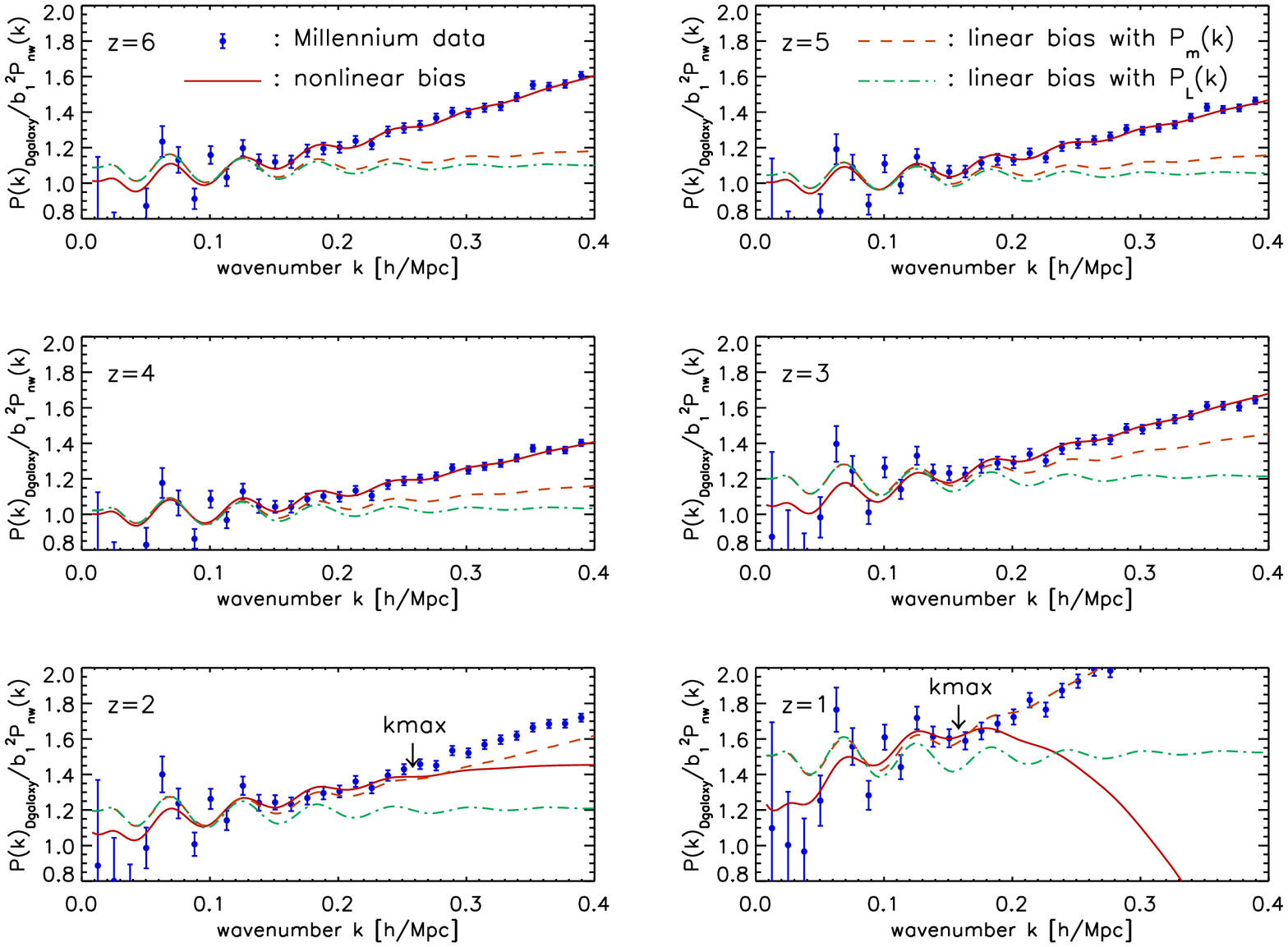}
}
\caption{
Same as Figure~\ref{fig9}, but for the Durham galaxy power spectrum
 (Dgalaxy). 
}%
\label{fig13}
\end{figure*}

Figures \ref{fig10} and \ref{fig11} show the galaxy power spectra
estimated from the MPA (Mgalaxy) and Durham (Dgalaxy) galaxy catalogues,
respectively. 
Here, we basically find the same story as we have found for the halo
power spectra (\S~\ref{sec:halo}): for $k<k_{max}$
 the non-linear bias model fits both galaxy power spectra (Mgalaxy and
 Dgalaxy), whereas the linear bias models fit neither.

The galaxy bias parameters extracted from Mgalaxy and Dgalaxy are summarized in
Table \ref{table:Mgbias} and \ref{table:Dgbias}, respectively. 
While the bias parameters are different for halo, Mgalaxy and Dgalaxy,
they follow the same trend: (i) $\tilde{b}_1$ becomes lower as the redshift
becomes lower, and (ii) $\tilde{b}_2$ also becomes lower as the redshift
becomes lower when $z>3$, but suddenly changes to large negative values at
$z\le 3$. As we have already pointed out in \S~\ref{sec:halo}, this
sudden peculiar change is most likely caused by the double-peak nature
of the likelihood function, owing  to the poor statistical power for
lower $k_{max}$ at lower $z$. In order to study $\tilde{b}_2$ further
with better statistics, one needs a bigger simulation. 

\subsubsection{Comparison with the simplest HOD predictions}
To give a rough theoretical guide for the galaxy bias parameters, 
we assume that each dark matter halo hosts one galaxy above a certain 
minimum mass.
This specifies the form of the HOD completely:
$\langle N|M\rangle=1$, with the same lower mass cut-off  as the
minimum mass of the halo, $M_{min}=1.72\times10^{10} M_\odot/h$.

This is utterly simplistic, and is probably not correct for describing
Mgalaxy or Dgalaxy. Nevertheless, we give the resulting values in 
Table \ref{table:Mgbias} and \ref{table:Dgbias}, which have been
computed from
\begin{equation}
b_i^{ST}=\frac
{\int_{M_{min}}^{M_{max}}\frac{dn}{dM}b_i(M)\langle N|M\rangle dM}
{\int_{M_{min}}^{M_{max}}\frac{dn}{dM}\langle N|M\rangle dM},
\end{equation}
where $dn/dM$ is the Sheth-Tormen mass function and $b_i(M)$ is the
$i$-th order bias parameter from \citet{scoccimarro/etal:2001}.
To compare with the non-linear bias parameters, we also calculate 
$\tilde{b}_2=b_2^{ST}/\tilde{b}_1$.

While these ``predictions'' give values that are reasonably close to the
ones obtained from the fits, they are many $\sigma$ away from the
best-fitting values. The freedom in the choice of the HOD may be used to
make the predicted values match the best-fitting values; however, such
an approach would require at least as many free parameters as the
non-linear bias parameters. Also, given that the {\it halo} bias
prediction fails to fit the halo power spectra, the HOD approach, which
is still based upon knowing the halo bias, is bound to fail as well.
 
\subsubsection{Effects on BAOs}
In Figures \ref{fig12} and \ref{fig13} we show 
how non-linear galaxy bias distorts the structure of BAOs.
Again, we find the same story as we have found for the halo bias:
the galaxy bias distorts BAOs more at higher $z$ because, for a given
mass, galaxies were rarer at higher redshifts and thus more highly
biased, while the quality of the fits is better at higher $z$ because of
less non-linearity in the matter clustering.

In all cases (halo, Mgalaxy and Dgalaxy) the non-linear bias model given
by Eq.~(\ref{eq:3rd_PT_Pk}) provides very good fits, and describes how
bias modifies BAOs.

\subsection{Mass dependence of  bias parameters and effects on
  BAOs}\label{sec:mass_dependence}
\begin{figure*}
\centering
\rotatebox{90}{
	\includegraphics[width=10cm]{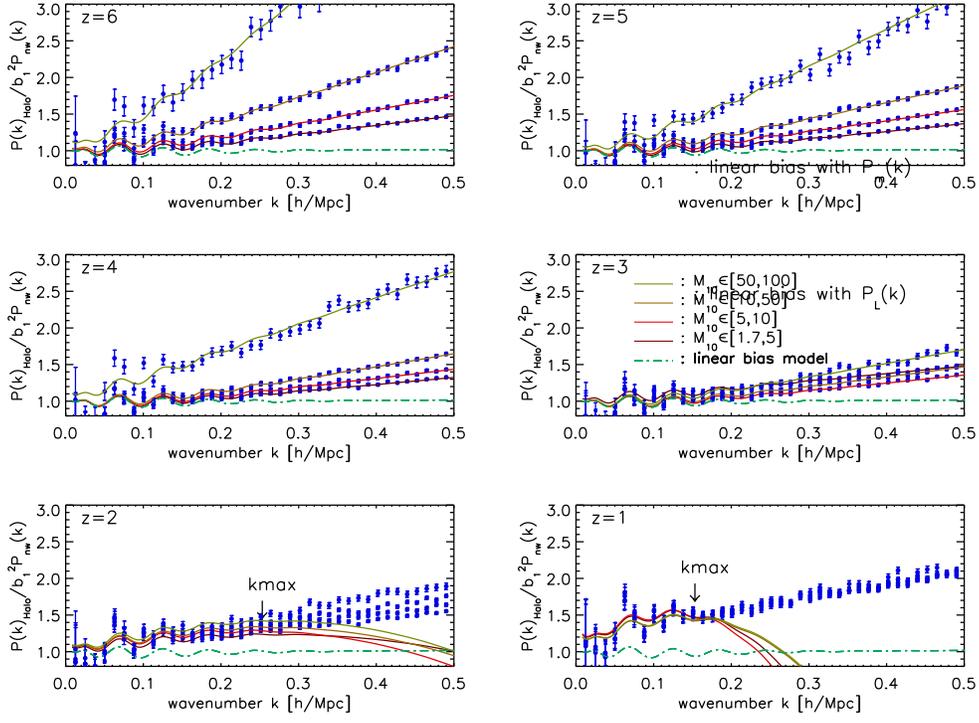}
}
\caption{
Mass dependence of distortion of BAOs due to non-linear bias.
Four mass bins, $M<5\times10^{10}M_\odot/h$, 
$5\times10^{10}M_\odot/h<M<10^{11}M_\odot/h$, 
$10^{11}M_\odot/h<M<5\times10^{11}M_\odot/h$, and 
$5\times10^{11}M_\odot/h<M<10^{12}M_\odot/h$, are shown.
($M_{10}$ stands for $M/(10^{10}M_\sun)$.)
All of the power spectra have been divided by a smooth power spectrum 
without baryonic oscillations from equation~(29) of \cite{eisenstein/hu:1998}. 
The errorbars show the Millennium Simulation data, while the solid 
lines show the PT calculation.
}%
\label{fig14}
\end{figure*}
\begin{figure*}
\centering
\rotatebox{90}{
	\includegraphics[width=10cm]{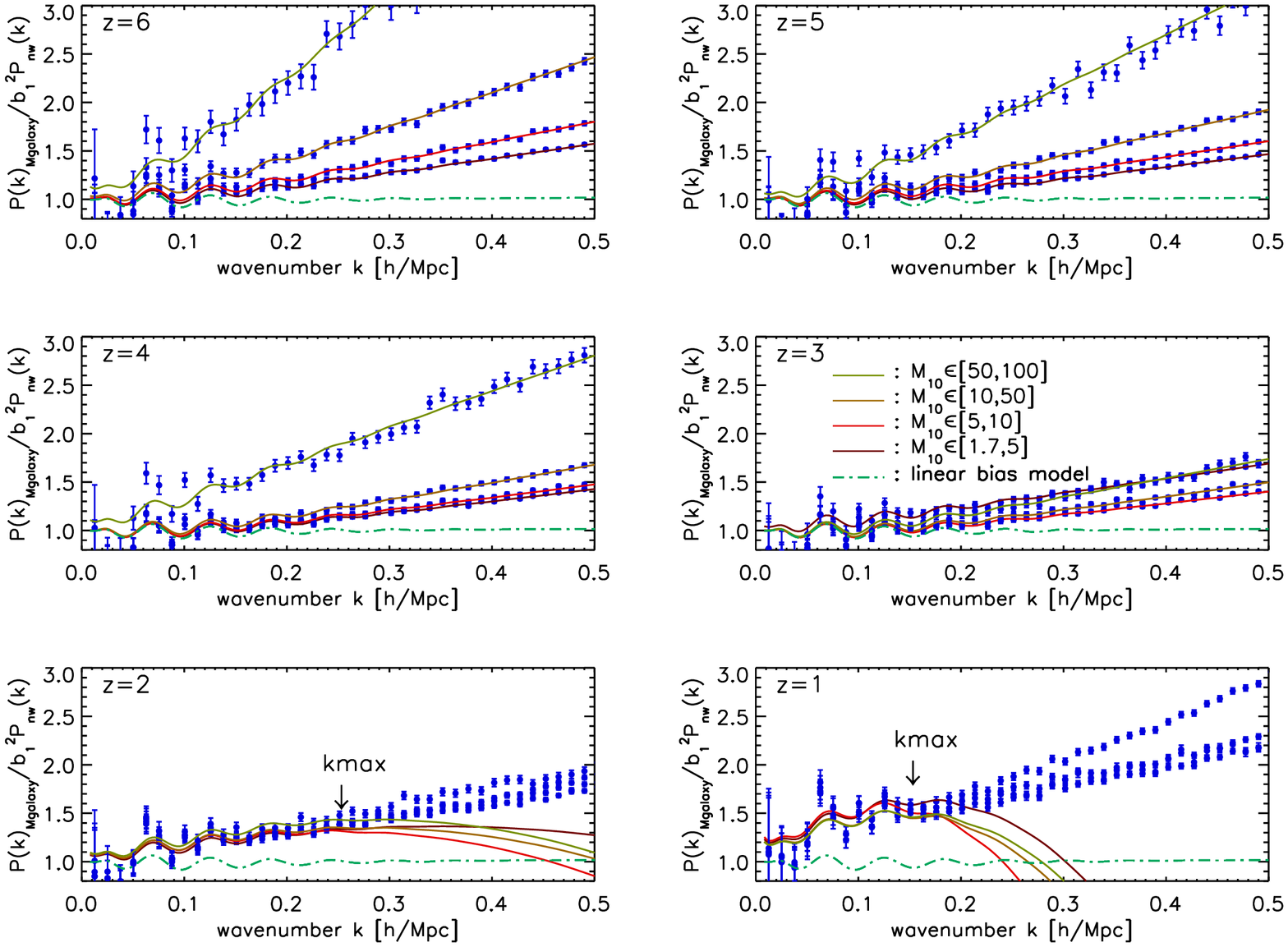}
}
\caption{
Same as Figure~\ref{fig14}, but for the MPA galaxy catalogue (Mgalaxy).
}%
\label{fig15}
\end{figure*}

\begin{deluxetable*}{cccccccc}




\tablecaption{Mass dependence of non-linear halo bias parameters (MPA halos)}
\tabletypesize{\footnotesize}

\tablenum{6}
\label{table:Mhbias_mdep}

\tablehead{\colhead{$z$} & \colhead{$\mathrm{M_{min}}$} & \colhead{$\mathrm{M_{max}}$} & \colhead{$\tilde{b}_1$} & \colhead{$\tilde{b}_2$} & \colhead{$P_0$} & \colhead{$b_1^{ST}$} & \colhead{$\tilde{b}_2^{ST}$} \\ 
\colhead{} & \colhead{($M_\odot/h$)} & \colhead{($M_\odot/h$)} & \colhead{} & \colhead{} & \colhead{($[h/\mathrm{Mpc}]^3$)} & \colhead{} & \colhead{} } 

\startdata
6 & 1.7E+10 & 5.0E+10 & 3.19$\pm$0.01 & 1.28$\pm$0.03 & 88.76$\pm$2.97 & 2.96 & 0.93 \\
 & 5.0E+10 & 1.0E+11 & 3.90$\pm$0.02 & 1.91$\pm$0.04 & 288.18$\pm$8.02 & 3.52 & 1.36 \\
 & 1.0E+11 & 5.0E+11 & 4.66$\pm$0.03 & 3.04$\pm$0.05 & 1029.19$\pm$18.84 & 4.41 & 2.28 \\
 & 5.0E+11 & 1.0E+12 & 6.41$\pm$0.14 & 5.76$\pm$0.21 & 6910.17$\pm$200.74 & 5.95 & 3.59 \\
\hline
5 & 1.7E+10 & 5.0E+10 & 2.55$\pm$0.01 & 0.71$\pm$0.03 & 31.51$\pm$2.06 & 2.41 & 0.48 \\
 & 5.0E+10 & 1.0E+11 & 3.09$\pm$0.01 & 1.19$\pm$0.04 & 120.84$\pm$5.69 & 2.84 & 0.81 \\
 & 1.0E+11 & 5.0E+11 & 3.78$\pm$0.02 & 1.79$\pm$0.04 & 402.11$\pm$12.40 & 3.55 & 1.48 \\
 & 5.0E+11 & 1.0E+12 & 5.14$\pm$0.07 & 3.55$\pm$0.11 & 2805.48$\pm$94.53 & 4.71 & 2.44 \\
\hline
4 & 1.7E+10 & 5.0E+10 & 2.08$\pm$0.01 & 0.38$\pm$0.04 & 10.90$\pm$1.19 & 2.01 & 0.15 \\
 & 5.0E+10 & 1.0E+11 & 2.51$\pm$0.01 & 0.66$\pm$0.04 & 42.34$\pm$3.52 & 2.33 & 0.40 \\
 & 1.0E+11 & 5.0E+11 & 3.05$\pm$0.01 & 1.08$\pm$0.04 & 161.22$\pm$8.11 & 2.90 & 0.92 \\
 & 5.0E+11 & 1.0E+12 & 3.80$\pm$0.05 & -4.08$\pm$0.09 & 3431.19$\pm$64.81 & 3.79 & 1.77 \\
\hline
3 & 1.7E+10 & 5.0E+10 & 1.39$\pm$0.01 & -1.83$\pm$0.05 & 241.59$\pm$9.58 & 1.47 & -0.25 \\
 & 5.0E+10 & 1.0E+11 & 1.75$\pm$0.01 & 0.11$\pm$0.05 & 2.48$\pm$0.29 & 1.67 & -0.10 \\
 & 1.0E+11 & 5.0E+11 & 2.09$\pm$0.01 & 0.35$\pm$0.04 & 20.95$\pm$3.22 & 2.04 & 0.19 \\
 & 5.0E+11 & 1.0E+12 & 2.78$\pm$0.02 & 0.82$\pm$0.06 & 171.31$\pm$21.03 & 2.57 & 0.60 \\
\hline
2 & 1.7E+10 & 5.0E+10 & 1.01$\pm$0.05 & -1.98$\pm$0.68 & 373.60$\pm$149.24 & 1.11 & -0.46 \\
 & 5.0E+10 & 1.0E+11 & 1.14$\pm$0.07 & -2.30$\pm$0.63 & 627.69$\pm$204.69 & 1.23 & -0.40 \\
 & 1.0E+11 & 5.0E+11 & 1.31$\pm$0.08 & -2.34$\pm$0.63 & 869.30$\pm$272.06 & 1.44 & -0.28 \\
 & 5.0E+11 & 1.0E+12 & 1.62$\pm$0.11 & -2.53$\pm$0.69 & 1566.40$\pm$476.07 & 1.75 & -0.05 \\
\hline
1 & 1.7E+10 & 5.0E+10 & 0.68$\pm$0.09 & -3.12$\pm$1.46 & 1315.40$\pm$447.14  & 0.86 & -0.58 \\
 & 5.0E+10 & 1.0E+11 & 0.75$\pm$0.10 & -3.24$\pm$1.46 & 1699.76$\pm$571.75 & 0.92 & -0.56 \\
 & 1.0E+11 & 5.0E+11 & 0.85$\pm$0.09 & -2.80$\pm$1.65 & 1783.07$\pm$683.57 & 1.02 & -0.53 \\
 & 5.0E+11 & 1.0E+12 & 0.99$\pm$0.11 & -2.82$\pm$1.95 & 2443.76$\pm$972.16 & 1.17 & -0.47
\enddata


\tablecomments{$z$: redshift\\
$\mathrm{M_{min}}$: minimum mass for a given bin\\
$\mathrm{M_{max}}$: maximum mass for a given bin\\
$\tilde{b}_1$, $\tilde{b}_2$, $P_0$: non-linear bias parameters\\
$b_1^{ST}$, $\tilde{b}_2^{ST}$: bias parameters from the Sheth-Tormen model, $\tilde{b}_2^{ST}$=$b_2^{ST}/\tilde{b}_1$
\\
\textit{
Caution: We estimate 1-$\sigma$ ranges for the low redshift ($z\le3$)
only for the peak which involves the maximum likelihood value. 
If two peaks in maginalized likelihood function are blended, we use
only unblended side of the peak to estimate the 1-$\sigma$ range.
}
}

\end{deluxetable*}

\begin{deluxetable*}{cccccccc}


\tabletypesize{\footnotesize}


\tablecaption{Mass dependence of non-linear galaxy bias parameters (MPA
 galaxies)} 

\tablenum{7}
\label{table:Mgbias_mdep}

\tablehead{\colhead{$z$} & \colhead{$\mathrm{M_{min}}$} & \colhead{$\mathrm{M_{max}}$} & \colhead{$\tilde{b}_1$} & \colhead{$\tilde{b}_2$} & \colhead{$P_0$} & \colhead{$b_1^{ST}$} & \colhead{$\tilde{b}_2^{ST}$} \\ 
\colhead{} & \colhead{($M_\odot/h$)} & \colhead{($M_\odot/h$)} & \colhead{} & \colhead{} & \colhead{($[h/\mathrm{Mpc}]^3$)} & \colhead{} & \colhead{} } 

\startdata
6 & 1.7E+10 & 5.0E+10 & 3.37$\pm$0.01 & 1.50$\pm$0.03 & 136.39$\pm$3.69 & 2.91 & 0.82 \\
 & 5.0E+10 & 1.0E+11 & 3.96$\pm$0.02 & 2.00$\pm$0.04 & 325.38$\pm$8.44 & 3.49 & 1.31 \\
 & 1.0E+11 & 5.0E+11 & 4.69$\pm$0.03 & 3.09$\pm$0.05 & 1078.72$\pm$19.25 & 4.23 & 2.01 \\
 & 5.0E+11 & 1.0E+12 & 6.43$\pm$0.14 & 5.79$\pm$0.20 & 7046.28$\pm$201.94 & 5.89 & 3.49 \\
\hline
5 & 1.7E+10 & 5.0E+10 & 2.77$\pm$0.01 & 0.93$\pm$0.03 & 63.17$\pm$2.99 & 2.38 & 0.40 \\
 & 5.0E+10 & 1.0E+11 & 3.16$\pm$0.01 & 1.27$\pm$0.04 & 144.20$\pm$6.16 & 2.82 & 0.77 \\
 & 1.0E+11 & 5.0E+11 & 3.81$\pm$0.02 & 1.84$\pm$0.04 & 432.51$\pm$12.80 & 3.41 & 1.28 \\
 & 5.0E+11 & 1.0E+12 & 5.15$\pm$0.07 & 3.60$\pm$0.11 & 2897.95$\pm$95.17 & 4.67 & 2.37 \\
\hline
4 & 1.7E+10 & 5.0E+10 & 2.33$\pm$0.01 & 0.58$\pm$0.03 & 32.25$\pm$2.25 & 1.98 & 0.11 \\
 & 5.0E+10 & 1.0E+11 & 2.59$\pm$0.01 & 0.74$\pm$0.04 & 56.91$\pm$4.08 & 2.32 & 0.37 \\
 & 1.0E+11 & 5.0E+11 & 3.09$\pm$0.02 & 1.13$\pm$0.04 & 179.81$\pm$8.52 & 2.79 & 0.77 \\
 & 5.0E+11 & 1.0E+12 & 3.83$\pm$0.05 & -4.09$\pm$0.09 & 3507.05$\pm$64.85 & 3.76 & 1.71 \\
\hline
3 & 1.7E+10 & 5.0E+10 & 1.62$\pm$0.01 & -2.07$\pm$0.05 & 431.79$\pm$12.04 & 1.45 & -0.22 \\
 & 5.0E+10 & 1.0E+11 & 1.84$\pm$0.01 & 0.19$\pm$0.04 & 7.04$\pm$1.04 & 1.66 & -0.10 \\
 & 1.0E+11 & 5.0E+11 & 2.14$\pm$0.01 & 0.38$\pm$0.04 & 27.64$\pm$3.67 & 1.96 & 0.12 \\
 & 5.0E+11 & 1.0E+12 & 2.80$\pm$0.02 & 0.84$\pm$0.06 & 191.24$\pm$21.65 & 2.55 & 0.57 \\
\hline
2 & 1.7E+10 & 5.0E+10 & 1.26$\pm$0.07 & -2.09$\pm$0.66 & 683.11$\pm$240.40 & 1.10 & -0.37 \\
 & 5.0E+10 & 1.0E+11 & 1.21$\pm$0.08 & -2.35$\pm$0.62 & 738.09$\pm$231.05 & 1.22 & -0.38 \\
 & 1.0E+11 & 5.0E+11 & 1.35$\pm$0.09 & -2.32$\pm$0.63 & 919.65$\pm$288.79 & 1.40 & -0.29 \\
 & 5.0E+11 & 1.0E+12 & 1.65$\pm$0.11 & -2.50$\pm$0.69 & 1602.17$\pm$480.23 & 1.74 & -0.06 \\
\hline
1 & 1.7E+10 & 5.0E+10 & 0.91$\pm$0.11 & -2.96$\pm$1.59 & 2344.13$\pm$802.55 & 0.86 & -0.43 \\
 & 5.0E+10 & 1.0E+11 & 0.79$\pm$0.11 & -3.28$\pm$1.51 & 1956.42$\pm$657.89 & 0.92 & -0.53 \\
 & 1.0E+11 & 5.0E+11 & 0.87$\pm$0.10 & -2.88$\pm$1.63 & 1964.64$\pm$738.71 & 1.00 & -0.51 \\
 & 5.0E+11 & 1.0E+12 & 1.01$\pm$0.11 & -2.80$\pm$2.01 & 2550.21$\pm$1007.23 & 1.16 & -0.46
\enddata


\tablecomments{$z$: redshift\\
$\mathrm{M_{min}}$: minimum mass for a given bin\\
$\mathrm{M_{max}}$: maximum mass for a given bin\\
$\tilde{b}_1$, $\tilde{b}_2$, $P_0$: non-linear bias parameters\\
$b_1^{ST}$, $\tilde{b}_2^{ST}$: bias parameters from the Sheth-Tormen model, $\tilde{b}_2^{ST}$=$b_2^{ST}/\tilde{b}_1$
\\
\textit{
Caution: We estimate 1-$\sigma$ ranges for the low redshift ($z\le3$)
only for the peak which involves the maximum likelihood value. 
If two peaks in maginalized likelihood function are blended, we use
only unblended side of the peak to estimate the 1-$\sigma$ range.
}
}

\end{deluxetable*}

So far, we have used all the available halos and galaxies in the
Millennium catalogues for computing the halo and galaxy power spectra.
In this section we divide the samples into different mass bins given by
$M<5\times10^{10}M_\odot/h$, 
$5\times10^{10}M_\odot/h<M<10^{11}M_\odot/h$, 
$10^{11}M_\odot/h<M<5\times10^{11}M_\odot/h$, 
$5\times10^{11}M_\odot/h<M<10^{12}M_\odot/h$,
and study how the derived bias parameters depend on mass.

The power spectra of the selected halos and galaxies in a given mass bin 
are calculated and fit in the exactly same manner as before.
Note that we shall use only the halo and Mgalaxy, as we expect 
that Dgalaxy would give similar results to Mgalaxy.

Figures \ref{fig14} and \ref{fig15}
show the results for the halo and galaxies, respectively.
To compare the power spectra of different mass bins in the same 
panel, and highlight the effects on BAOs at the same time, 
we have divided the power spectra by a non-oscillating matter power 
spectrum from equation (29) of \citet{eisenstein/hu:1998} with 
the best-fitting $\tilde{b}_1^2$ from each mass bin multiplied.
These figures show the expected results: the larger the mass is, the
larger the non-linear bias becomes. Nevertheless, the 3rd-order PT
calculation captures the dependence on mass well, and there is no
evidence for failure of the PT for highly biased objects. 

In Tables \ref{table:Mhbias_mdep} and \ref{table:Mgbias_mdep} we give
values of the measured bias parameters as well as the ``predicted'' values.
For all redshifts we see the expected trend again: the higher the mass
is,  the larger the effective linear bias ($\tilde{b}_1$) is.
The same is true for $\tilde{b}_2$ for $z>3$, 
while it is not as apparent for lower redshifts, 
and eventually becomes almost fuzzy for $z=1$. Again, these are probably
due to the lack of statistics due to lower values of $k_{max}$ at lower
$z$, and we need a bigger simulation to handle these cases with more
statistics. 

The high values of bias do not mean failure of PT. The PT galaxy power
spectrum model fails only when $\Delta^2_m(k,z)$ exceeds $\sim 0.4$ (Paper
I), or the locality of bias is violated. 
Overall, we find that the non-linear bias model given by
Eq.~(\ref{eq:3rd_PT_Pk}) performs well for halos and galaxies with 
all mass bins, provided that we use the data only up to $k_{max}$
determined from the matter power spectra. This implies that the locality
assumption is a good approximation for $k<k_{max}$; however, is it good
enough for us to extract cosmology from the observed galaxy power
spectra?

\section{Cosmological parameter estimation with the non-linear bias
 model}
\label{sec:bias}
In the previous sections we have shown that the 3rd-order PT galaxy
power spectrum given by
Eq.~(\ref{eq:3rd_PT_Pk}) provides good fits to the galaxy power spectrum
data from the Millennium Simulation.

However, we must not forget that Eq.~(\ref{eq:3rd_PT_Pk}) contains {\it
3 free parameters}, $\tilde{b}_1$, $\tilde{b}_2$, and $P_0$. With 3
parameters it may seem that it should not be so difficult to fit smooth
curves like those shown in, e.g., Figure~\ref{fig10}.

While the quality of fits is important, it is not the end of story. We
must also show that Eq.~(\ref{eq:3rd_PT_Pk}) can be used for extracting the
{\it correct cosmological parameters} from the observed galaxy power
spectra. 

In this section we shall extract the distance scale from the
galaxy power spectra of the Millennium Simulation, and compare them
with the input values that were used to generate the simulation. 
If they do not agree, Eq.~(\ref{eq:3rd_PT_Pk}) must be discarded. If
they do, we should proceed to the next level by including non-linear
redshift space distortion. 

\subsection{Measuring Distance Scale}
\label{sec:da}
\subsubsection{Background}
Dark energy influences the expansion rate of the universe as well as the
growth of structure \citep[see][for a recent
review]{copeland/sami/tsujikawa:2006}. 

The cosmological distances, such as the luminosity distance, $D_L(z)$, and angular
diameter distance, $D_A(z)$, are powerful tools for measuring the expansion rates
of the universe, $H(z)$, over a wide range of redshifts.
Indeed, it was $D_L(z)$ measured out to high-$z$ ($z\le
1.7$) Type Ia
supernovae that gave rise to the first compelling evidence for the
existence of dark energy \citep{riess/etal:1998,perlmutter/etal:1999}.
The CMB power spectrum provides us with a high-precision measurement of $D_A(z_*)$
out to the photon decoupling epoch,
$z_*\simeq 1090$ \citep[see][for the latest determination from the WMAP
5-year data]{komatsu/etal:prep}.

The galaxy power spectrum can be used for measuring $D_A(z)$ 
 as well as $H(z)$ over a wider
range of redshifts. From galaxy surveys we find  three-dimensional
positions of galaxies by measuring their angular positions on the sky as
well as their redshifts. We can then estimate the two-point correlation
function of galaxies as a function of the angular separation,
$\Delta\theta$, and the redshift separation, $\Delta z$. 
To convert $\Delta\theta$ and $\Delta z$ into the comoving separations
perpendicular to the line of sight, $\Delta r_\perp$, and those along the line
of sight, $\Delta r_\parallel$, one needs to know $D_A(z)$ and $H(z)$,
 respectively, as
\begin{eqnarray}
  \Delta r_\perp &=& (1+z)D_A(z)\Delta\theta,\\
  \Delta r_\parallel &=& \frac{c\Delta z}{H(z)},
\end{eqnarray}
where $(1+z)$ appears because $D_A(z)$ is the proper (physical) angular
diameter distance, whereas $\Delta r_\perp$ is the comoving separation. 
Therefore, if we know $\Delta r_\perp$ and $\Delta r_\parallel$ {\it a
priori}, then we may use the above equations to measure $D_A(z)$ and
$H(z)$.

The galaxy power spectra contain at least three distance scales which
may be used in the place of $\Delta r_\perp$ and $\Delta r_\parallel$:
(i) the sound horizon size at the so-called baryon drag epoch,
$z_{drag}\simeq 1020$, at which
baryons were released from the baryon-photon plasma, (ii) the photon
horizon size at the matter-radiation equality, $z_{eq}\simeq 3200$, and
(iii) the Silk damping scale \citep[see, e.g.,][]{eisenstein/hu:1998}.

In Fourier space, we may write the observed power spectrum as
\citep{seo/eisenstein:2003}
\begin{eqnarray}\label{eq:pkDaHz}
P_{obs}&(&k_\parallel,k_\perp,z)
=
\left(\frac{D_A(z)}{D_{A,\mathrm{true}}(z)}\right)^2
\left(\frac{H_{\mathrm{true}}(z)}{H(z)}\right)
\nonumber
\\
&\times&
P_{\mathrm{true}}
\left(
\frac{D_{A,\mathrm{true}}(z)}{D_A(z)}k_\perp,
\frac{H(z)}{H_{\mathrm{true}}(z)}k_\parallel,z
\right),
\end{eqnarray}
where $k_\perp$ and $k_\parallel$ are the wavenumbers perpendicular to and
parallel to the line of sight, respectively, and
$P_{true}(k)$, $D_{A,true}(z)$, and $H_{true}(z)$ are the true,
underlying values. We then vary $D_A(z)$ and $H(z)$, trying to estimate
$D_{A,true}(z)$ and $H_{true}(z)$.

There are two ways of measuring $D_A(z)$ and $H(z)$
 from the galaxy power spectra:
\begin{itemize}
\item[(1)] Use BAOs. The BAOs contain the information of one of the
	   standard rulers, the sound horizon size at $z_{drag}$. This
	   method relies on measuring only the phases of BAOs, which are
	   markedly insensitive to all the non-linear
	   effects (clustering, bias, and redshift space distortion)
	   \citep{seo/eisenstein:2005,eisenstein/seo:2007,nishimichi/etal:2007,
      smith/scoccimarro/sheth:2008,angulo/etal:2008,sanchez/baugh/angulo:prep,
      seo/etal:prep,shoji/jeong/komatsu:prep}, 
      despite the fact that the amplitude is
	   distorted by non-linearities (see Figures~\ref{fig4},
	   \ref{fig9}, \ref{fig12}, and
	   \ref{fig13}). Therefore, BAOs provide a robust means to
	   measure $D_A(z)$ and $H(z)$, and they have been used for
	   determining $D_A^2H^{-1}$ out to $z=0.2$ from the SDSS main
	   galaxy sample and 2dFGRS, as well as to $z=0.35$ from the
	   SDSS Luminous Red Galaxy (LRG) sample
	   \citep{eisenstein/etal:2005,percival/etal:2007c}; however,
	   since they use only one 
	   standard ruler, the constraints on $D_A(z)$ and $H(z)$ from
	   the BAO-only analysis are weaker than the full analysis
	   \citep{shoji/jeong/komatsu:prep}. 
\item[(2)] Use the \textit{entire} shpae of the power spectrum. This
	   approach gives the best determination (i.e., the smallest
	   error) of $D_A(z)$ and $H(z)$, as it uses all the standard
	   rulers encoded in the galaxy power spectrum; however, one must
	   understand the distortions of the shape of the power spectrum
	   due to non-linear effects. The question is, ``is the
	   3rd-order (or higher) PT good enough for correcting the
	   key non-linear effects?'' 
\end{itemize}
In this paper we show, for the first time, that we can extract 
the distance scale 
using the 3rd-order PT galaxy power spectrum in real space. While we have not
yet included the effects of redshift space distortion, this is a
significant step towards extracting $D_A(z)$ and $H(z)$ from the entire
shape of the power spectrum of galaxies.
We shall address the effect of non-linear redshift space distortion in
the future work.

\subsubsection{Method: Measuring ``Box Size'' of the Millennium Simulation}
In real space simulations (as opposed to redshift space ones), 
there is only one distance scale in the problem: 
the box size of the simulation, $L_{\mathrm{box}}$, which
is $L_{\mathrm{box}}^{\mathrm{(true)}}=500~\mathrm{Mpc}/h$
for the Millennium simulation.
Then, ``estimating  
the distance scale from the Millennium Simulation'' becomes
equivalent to ``estimating $L_{\rm box}$ from the Millennium
Simulation.
Eq.~(\ref{eq:pkDaHz}) now leads:
\begin{equation}\label{eq:pkLbox}
P_{obs}(k,L_{\mathrm{box}})
=
\left(
\frac{L_{\mathrm{box}}}{L_{\mathrm{box}}^{\mathrm{(true)}}}
\right)^3
P_{\mathrm{true}}
\left(
\frac{L_{\mathrm{box}}^{\mathrm{(true)}}}{L_{\mathrm{box}}}k
\right).
\end{equation}

As we estimate the variance of power spectrum from the
observed power spectrum, 
we need to rescale the variance 
when the normalization of the observed power spectrum changes :
\begin{equation}
\label{eq:sigpkLbox}
\sigma_{Pi}^2(L_{\mathrm{box}})=
\left(
\frac{L_{\mathrm{box}}}{L_{\mathrm{box}}^{\mathrm{(true)}}}
\right)^6
\sigma_{Pi}^2(L_{\mathrm{box}}^{\mathrm{(true)}})
\end{equation}

We estimate $L_{\mathrm{box}}$ using 
the likelihood function given by
\begin{eqnarray}
\label{eq:LLbox}
\mathcal{L}(\tilde{b}_1,\tilde{b}_2,&P_0&,L_{\mathrm{box}})
=\prod_{k_i<k_{max}}
\frac{1}{\sqrt{2\pi\sigma_{Pi}^2(L_{\mathrm{box}})}}
\nonumber
\\
&\times&
\exp
\left[-
\frac{
\left\{
P_{obs}(k_i/\alpha)
-
P_{g}(k_i/\alpha)/\alpha^3
\right\}^2
}{2\sigma_{P}^2(k_i/\alpha)}
\right],
\end{eqnarray}
where 
$\alpha=L_{\mathrm{box}}/L_{\mathrm{box}}^{\mathrm{(true)}}$.

The likelihood function, Eq.~(\ref{eq:LLbox}), still depends upon the
bias parameters that we wish to eliminate. Therefore we marginalize 
the likelihood function over all the bias parameters with flat
priors.\footnote{Note that this is the most conservative analysis one can do. In reality we can use the
bispectrum for measuring $\tilde{b}_1$ and $\tilde{b}_2$, which would
give appropriate priors on them (see \S~\ref{sec:bispectrum}). We shall
report on the results from this analysis elsewhere.} 
We obtain (see also Appendix~\ref{sec:appB}):
\begin{equation}
\mathcal{L}(L_{\mathrm{box}})
=
\int_0^\infty d\tilde{b}_1^2
\int_{-\infty}^\infty d\tilde{b}_2
\int_{-\infty}^\infty dP_0~
\mathcal{L}(\tilde{b}_1,\tilde{b}_2,P_0,L_{\mathrm{box}}).
\end{equation}

Hereafter, we shall simply call $L_{\mathrm{box}}$ as $D$ for 
`distance scale'. $D$ is closely related to the angular diameter distance,
$D_A(z)$, and the expansion rate, $H(z)$, in real surveys. (See, \S 5.1.1)

\subsubsection{Results: Unbiased Extraction of the distance scale  from the Millennium Simulation}
\begin{figure*}
\centering
\rotatebox{90}{
	\includegraphics[width=10cm]{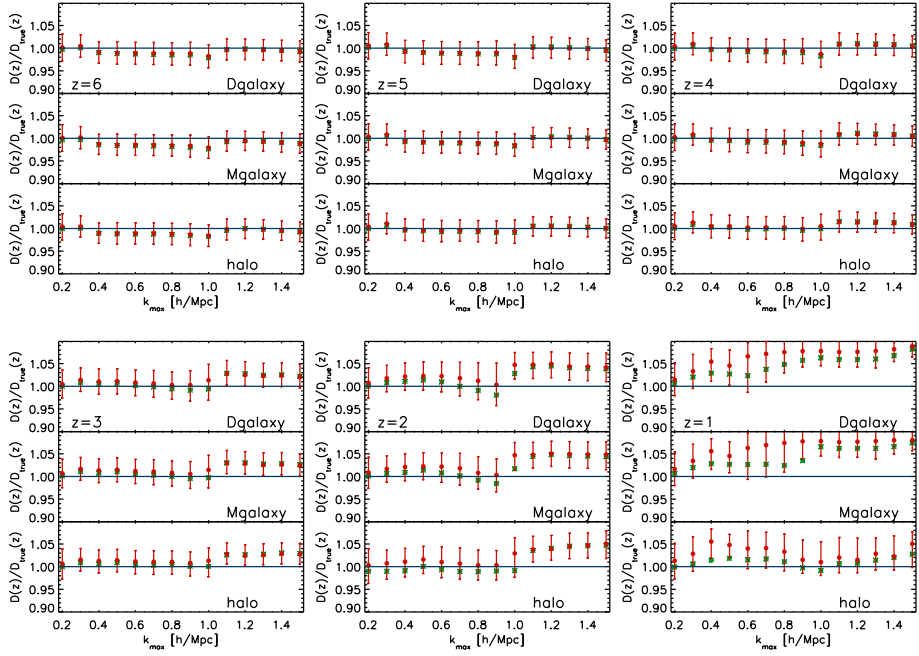}
}
\caption{
 Distance scale extracted from the Millennium Simulation
 using the 3rd-order PT galaxy power spectrum given by
 Eq.~(\ref{eq:3rd_PT_Pk}), divided by the true value. 
The mean of the likelihood ({\it stars}), and the maximum likelihood values
 ({\it filled circles}) and  the corresponding
 1-$\sigma$ intervals ({\it errorbars}), are shown as a function of 
 maximum wavenumbers used in the fits, $k_{max}$. 
We find 
$D/D_{true}=1$ 
to within the 1-$\sigma$ errors from all the
 halo/galaxy catalogues (``halo,'' ``Mgalaxy,'' and ``Dgalaxy'') at all
 redshifts, provided that we use 
$k_{max}$ estimated from the matter power spectra, 
$k_{max}=0.15$, 0.25, 1.0, 1.2, 1.3, and 1.5 at $z=1$, 2, 3, 4, 5, and
 6, respectively (see Table~\ref{table:kmax}).
 Note that the errors on $D$ do not decrease as $k_{max}$ increases
 due to degeneracy between $D$ and the bias parameters. 
 See Figure (\ref{fig18}) 
 and (\ref{fig19}) for further analysis.
}%
\label{fig16}
\end{figure*}
\begin{figure*}
\centering
\rotatebox{90}{
	\includegraphics[width=10cm]{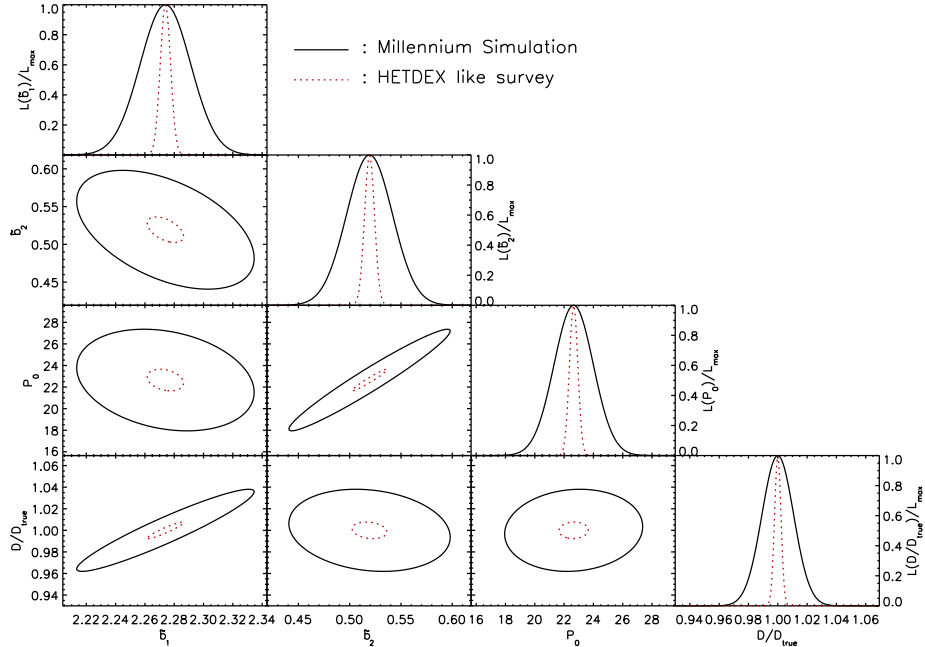}
}
\caption{
Same as Figure \ref{fig8}, but including the
distance scale $D/D_{true}$.
}%
\label{fig17}
\end{figure*}

In Figure \ref{fig16} we show 
$D(z)/D_{true}(z)$
estimated from the halo,
Mgalaxy, and Dgalaxy catalogues at $z=1$, 2, 3, 4, 5, and 6.
The maximum likelihood values (filled circles) and the corresponding
1-$\sigma$ intervals (errorbars), as well as the mean of the likelihood
({\it stars}) are shown. 
We find 
$D(z)/D_{true}(z)=1$ 
to within the 1-$\sigma$ errors from
{\it all of the halo/galaxy catalogues} at {\it all redshifts}, 
provided that we use $P_{obs}(k)$ only up to $k_{max}$ that has been
determined unambiguously from the matter power spectrum (see
Table~\ref{table:kmax}). 
Not only does this provide a strong support for the validity of
Eq.~(\ref{eq:3rd_PT_Pk}), but also it provides a practical means for
extracting 
$D$ from the full shape of the observed galaxy power spectra.

Despite a small volume of the Millennium Simulation and the use of flat
priors on the bias parameters upon marginalization, we could determine 
$D$ to about 2.5\% accuracy. 

In addition, we also find that the error on $D$ hardly decreases 
even though $k_{max}$ increases. 
It is because of the degeneracy between $D$ and the bias parameters.
In order to see how strongly degenerate they are,
we calculate correlations between pairs of parameters
($\tilde{b}_1$,$\tilde{b}_2$,$P_0$,$D/D_{true}$)
by the Fisher information matrix from Eq. (\ref{eq:fisher_bias}).

Figure \ref{fig17} shows both one-dimensional
marginalized constraints and two-dimensional joint marginalized 
constraints of 2-$\sigma$ range ($95.45\%$ CL) for 
the bias parameters and the distance scale.
This figure indicates that when we include the distance scale, the
correlations between bias parameters become milder. It is mainly
due to the correlation between the distance scale and $\tilde{b}_1$
making the constraint on $\tilde{b}_1$ weaker.
On the other hand, the one-dimensional marginalized 
likelihood functions for $\tilde{b}_2$ and $P_0$ are hardly changed.
The remaining degeneracies are those between ($\tilde{b}_2$,$P_0$) and
($\tilde{b}_1$,$D/D_{true}$). 
These degeneracies would be broken when we
include the information from the bispectrum, as the bispectrum
will measure $\tilde{b}_1$ and $\tilde{b}_2$.

\subsubsection{Optimal estimation of the distance scale} 
\begin{figure*}
\centering
\rotatebox{90}{
	\includegraphics[width=10cm]{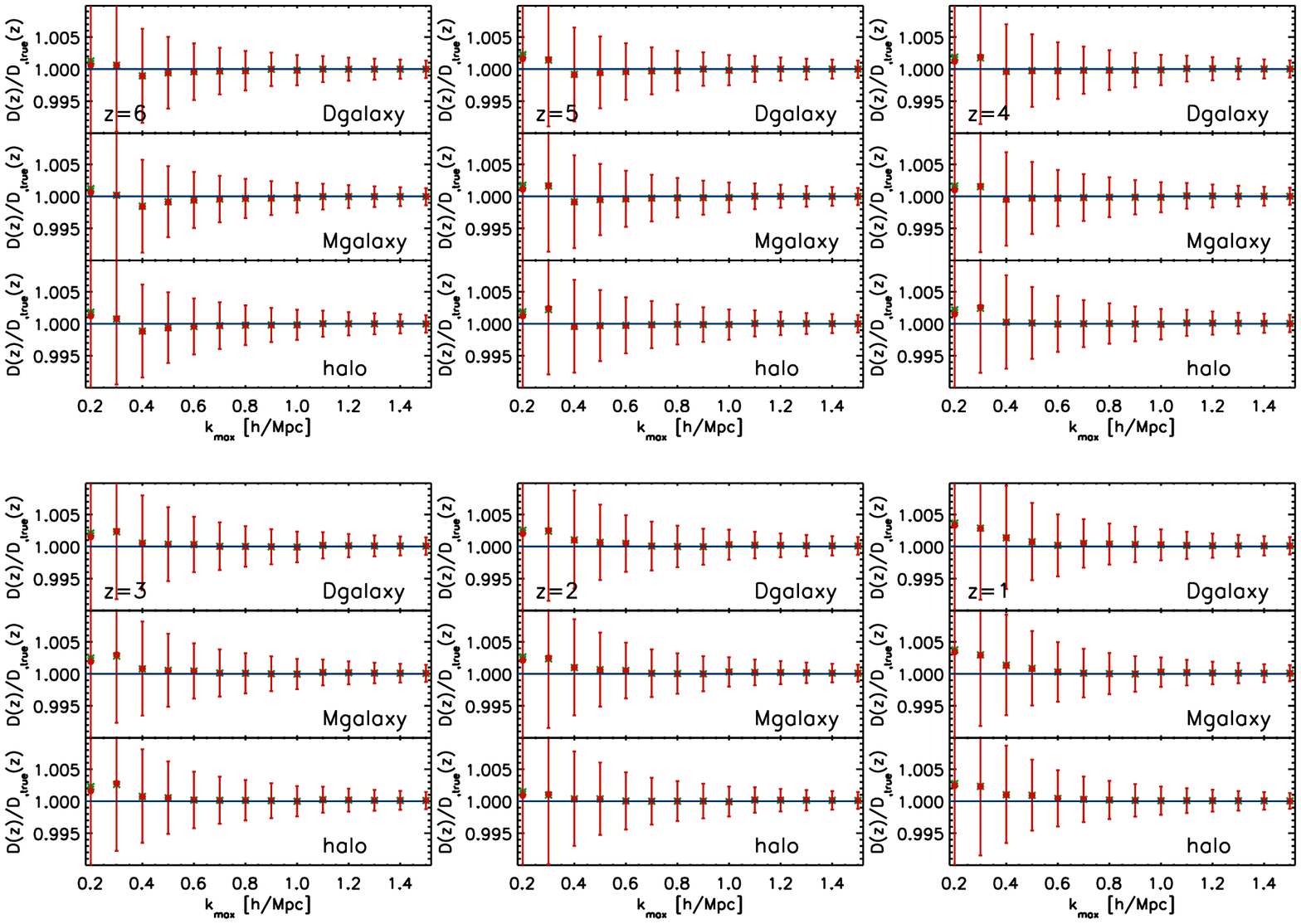}
}
\caption{
Same as Figure \ref{fig16}, but  
with $\tilde{b}_1$ and $\tilde{b}_2$ fixed at the best-fitting values.
The 1-$\sigma$ ranges for $D$ are
$1.5 \%$ 
and 
$0.15 \%$ 
for $k_{max}=0.2~h/\mathrm{Mpc}$ and $k_{max}=1.5~h/\mathrm{Mpc}$,
respectively. 
The errors on $D$ decrease as $k_{max}$ increases, but the scaling is 
still milder than $1/\sqrt{\sum_{k<k_{max}} N_k}$.
}%
\label{fig18}
\end{figure*}
\begin{figure*}
\centering
\rotatebox{90}{
	\includegraphics[width=10cm]{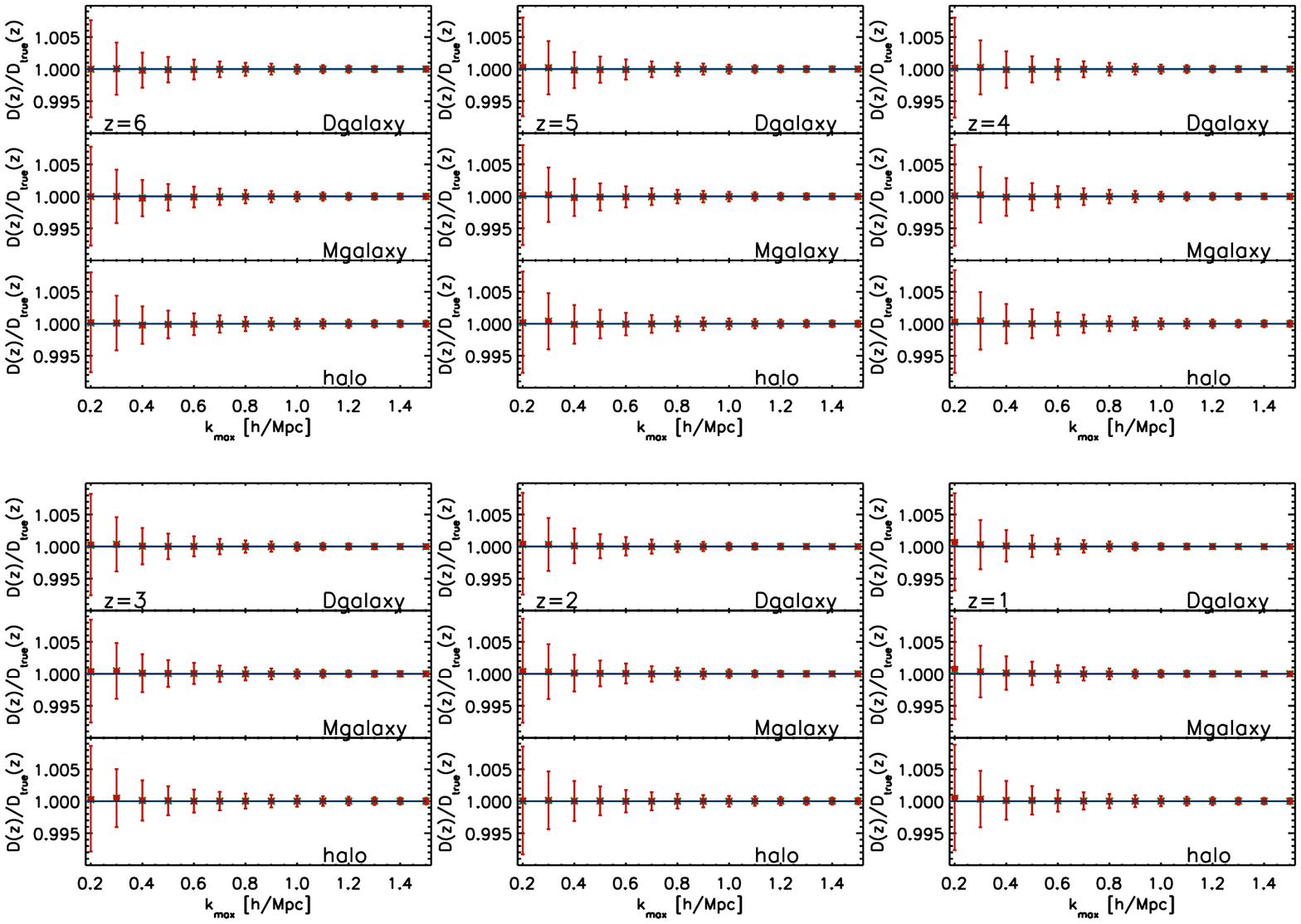}
}
\caption{
Same as Figure \ref{fig16}, but  
with $\tilde{b}_1$, $\tilde{b}_2$ and $P_0$ fixed at the best-fitting values.
The 1-$\sigma$ ranges for $D$ are
$0.8 \%$ 
and 
$0.05 \%$ 
for $k_{max}=0.2~h/\mathrm{Mpc}$ and $k_{max}=1.5~h/\mathrm{Mpc}$,
respectively. 
The errors on $D$ decrease as $k_{max}$ increases as 
$1/\sqrt{\sum_{k<k_{max}} N_k}$.
}%
\label{fig19}
\end{figure*}
The constraint we find from the previous subsection 
will get better when we include the bispectrum,
as the reduced bispectrum provides independent and strong 
constraints on $\tilde{b}_1$ and $\tilde{b}_2$
\citep{sefusatti/etal:2006}.

How much will it be better?
First, let us assume that we know the exact values of 
$\tilde{b}_1$ and $\tilde{b}_2$.
In this case, we get the error on $D$ by marginalizing only over $P_0$
while setting $\tilde{b}_1$ and $\tilde{b}_2$ to be the best-fitting values, 
i.e.
\begin{equation}\label{eq:Lalpha_b1b2_fixed}
\mathcal{L}^{\mathrm{fix}~\tilde{b}_1\tilde{b}_2}(D)
=\int_{-\infty}^{\infty} dP_0 
\mathcal{L}
(
\tilde{b}_1^{\mathrm{bf}},
\tilde{b}_2^{\mathrm{bf}},
P_0,D
)
\end{equation}
where $\tilde{b}_1^{\mathrm{bf}}$ and $\tilde{b}_2^{\mathrm{bf}}$
denote the best-fitting values of $\tilde{b}_1$ and $\tilde{b}_2$ for
each $k_{max}$, respectively.
In Figure \ref{fig18}, we show $D/D_{true}$ estimated
from Eq. (\ref{eq:Lalpha_b1b2_fixed}).
This figure shows that we can extract $D$ to about $1.5\%$ accuracy even
for the low $k_{max}=0.2~h/\mathrm{Mpc}$, and the error decreases further 
to $0.15\%$ for $k_{max}=1.5~h/\mathrm{Mpc}$.
Note that the uncertainties on $D/D_{true}$ decrease as $k_{max}$ increases
as expected. The reason is because 
fixing $\tilde{b}_1$ and $\tilde{b}_2$ 
breaks the degeneracy between them and the distance scale.

In reality, the bias parameters estimated from the bispectrum 
have finite errors, and thus the accuracy of extracting $D$ will be 
somewhere in between Figure \ref{fig16} 
and Figure \ref{fig18}.
The result of the full analysis including both power spectrum and
bispectrum of Millennium Simulation will be reported elsewhere.

In the ideal situation where we completely understand the complicated 
halo/galaxy formation, we may be able to calculate the 
three bias parameters from the first principle.
This \textit{ideal} determination of bias parameters will provide
more accurate constraints on the distance scale $D$. 
In this case, we get the likelihood function
by fixing all the bias parameters to their best-fitting values :
\begin{equation}\label{eq:Lalpha_bias_fixed}
\mathcal{L}^{\mathrm{fix~bias}}(D)
=
\mathcal{L}
(
\tilde{b}_1^{\mathrm{bf}},
\tilde{b}_2^{\mathrm{bf}},
P_0^{\mathrm{bf}},D
)
\end{equation}
By knowing all the bias parameters, we can extract the distance
scale $D$ to $0.8\%$ accuracy for $k_{max}=0.2~h/\mathrm{Mpc}$. 
The error decreases further to $0.05\%$ for $k_{max}=1.5~h/\mathrm{Mpc}$.
(See Figure (\ref{fig19}))

\subsubsection{Forecast for a HETDEX-like survey}
\begin{figure}
\centering
\rotatebox{90}{
	\includegraphics[width=6.5cm]{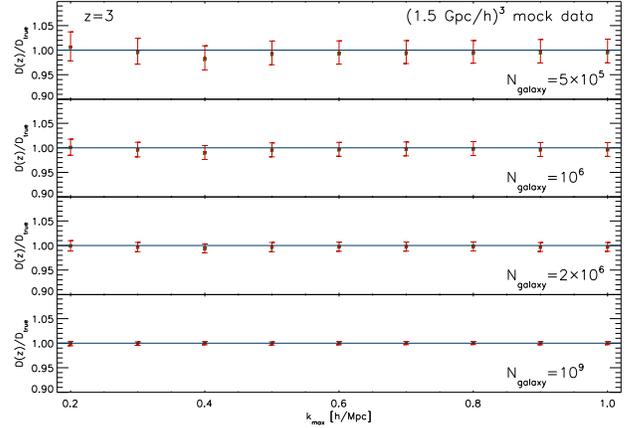}
}
\caption{
Projected constraints on 
 $D$ 
 at $z=3$ from a HETDEX-like survey with 
 the survey volume of $(1.5~\mathrm{Gpc/h})^3$. We have used the
 best-fitting 3rd-order PT power spectrum of MPA halos in the Millennium
 Simulation for generating a mock simulation
 data. We show the results for the number of objects of
 $N_{galaxy}=2\times 10^5$, $10^6$, $2\times 10^6$, and $10^9$, from the
 top  to bottom panels, respectively, for which we find the projected 1-$\sigma$
 errors of 2.5\%, 1.5\%, 1\%, and 0.3\%, respectively.
}%
\label{fig20}
\end{figure}

The planned future surveys would cover a larger volume than the
Millennium Simulation. Also, since the real surveys would be limited by
their continuum/flux sensitivity, they would not be able to detect 
all galaxies that were resolved in the Millennium Simulation. 
In this subsection we explore how the constraints would be 
affected by the volume and the number of objects.

To simulate the mock data, we take a simplified approach: we take 
our best-fitting  power spectrum at $z=3$, i.e.,
Eq.~(\ref{eq:3rd_PT_Pk}) fit to the  power spectrum of MPA halos in the 
Millennium Simulation at $z=3$, and add random Gaussian noise to it with
the standard deviation given by Eq.~(\ref{eq:varpk}).
To compute the standard deviation we need to specify the survey volume,
which determines the fundamental wavenumber, $\Delta k$, as $\Delta
k=2\pi/V_{survey}^{1/3}$. We use the volume that would be surveyed by
the HETDEX survey \citep{hill/etal:2004},
$V_{survey}=(1.5~\mathrm{Gpc}/h)^3$, which is 27 times as large as the
volume of the Millennium Simulation. We then vary the number of galaxies,
$N_{galaxy}$, which determines the shot noise as
$P_{shot}=1/n=V_{survey}/N_{galaxy}$. 
We have generated only one realization, and repeated the same analysis
as before to extract $D_A$ from the mock HETDEX data.

In Figure \ref{fig20} we show $D/D_{true}$ as a
function of $k_{max}$ and $N_g$. For $N_{galaxy}=10^9$, which gives the
same number density as the Millennium Simulation, the projected 
error on $D$ is 0.3\%, or 8 times better than the original result
presented in Figure~\ref{fig16}.
Since the volume is 27 times bigger, the statistics alone would reduce
the error by a factor of about 5.

The other factor of about 1.5 comes from the fact that the variance 
of the distance scale estimated from the Millennium Simulation 
lies on the tail of the distribution of the variance of the 
distance scale, (See, appendix C)
while the error estimated from the HETDEX volume mock is close to the peak of
PDF of the variance.

However, real surveys will not get as high the number density as the
Millennium Simulation. For example, the HETDEX survey will detect about
one million Ly$\alpha$ emitting galaxies, i.e., $N_{galaxy}=10^6$. 
In Figure~\ref{fig20} we show that 
the errors on $D$ increase from 0.3\% for $N_{galaxy}=10^9$ to 
1\%, 1.5\%, and 3\% for $N_{galaxy}=2\times 10^6$, $10^6$, and $2\times 10^5$, respectively.

Finally, we note that these forecasts are not yet final, as we have not
included the effect of non-linear redshift space distortion. Also,
eventually one needs to repeat this analysis using the ``super Millennium
Simulation'' with a bigger volume. 

\section{Discussion and Conclusions}
\label{sec:conclusion}
Two main new results that we have presented in this paper are:
\begin{itemize}
 \item The 3rd-order PT galaxy power spectrum given by
       Eq.~(\ref{eq:3rd_PT_Pk}), which is based upon the assumption that
       the number density of galaxies at a given location is a local
       function of the underlying matter density at the same location
       \citep{fry/gaztanaga:1993} plus stochastic noise
       \citep{mcdonald:2006}, fits the halo as well as galaxy power spectra
       estimated from the Millennium Simulation at high
       redshifts, $1\le z\le 6$, up to the maximum wavenumber,
       $k_{max}$, that has been determined from the matter power spectrum.
 \item When 3 galaxy bias parameters, $\tilde{b}_1$, $\tilde{b}_2$, and
       $P_0$, are marginalized over, the 
       3rd-order PT galaxy power spectrum fit to  the Millennium
       Simulation yields the correct (unbiased)
       distance scale to within the statistical error of the simulation,
       $\sim 3\%$.
\end{itemize}
These results suggest that the 3rd-order PT provides us with a practical
means to extract the cosmological information from the observed galaxy
power spectra at high redshifts, i.e., $z>1$, accurately.

We would like to emphasize that our approach does not require
simulations to calibrate the model. The 3rd-order PT is based upon the
solid physical framework, and the only assumption made for the galaxy
formation is that it is a local process, at least on the scales where
the 3rd-order PT is valid, i.e., $k<k_{max}$.
The only serious drawback so far is that the 3rd-order PT breaks down at
low redshifts, and thus it cannot be applied to the current generation
of survey data such as 2dFGRS and SDSS. However, the planned future
high-$z$ surveys would benefit immensely from the PT approach.

The practical application of our approach may proceed as follows:
\begin{itemize}
 \item [(1)] Measure the galaxy power spectra at various
       redshifts. When we have $N$ redshift bins, the number of bias
       parameters is $3N$, as the bias parameters evolve with $z$.
 \item [(2)] Calculate $k_{max}(z)$ from the condition,
       $\Delta_m^2(k_{max},z)=0.4$, where
       $\Delta_m^2(k,z)=k^3P_{\delta\delta}(k,z)/(2\pi^2)$ 
       is computed from the fiducial cosmology, e.g., the WMAP 5-year
       best-fitting parameters. The results should not be sensitive to
       the exact values of $k_{max}(z)$.
 \item [(3)] Fit Eq.~(\ref{eq:3rd_PT_Pk}) to the observed galaxy spectra
       up to $k_{max}(z)$
       at all $z$ simultaneously for extracting the cosmological parameters.
\end{itemize}
In addition to this, we should be able to improve upon the accuracy of
parameter determinations by including the bispectrum as well, as the
bispectrum basically fixes $\tilde{b}_1$ and $\tilde{b}_2$
\citep{sefusatti/komatsu:2007}. 
Therefore, the step (3) may be replaced by
\begin{itemize}
\item [(3')] Fit Eq.~(\ref{eq:3rd_PT_Pk}) to the observed galaxy
      spectra        up to $k_{max}(z)$, and fit the PT bispectrum to the
      observed galaxy bispectra up to the same $k_{max}(z)$, 
       at all $z$ simultaneously for extracting the cosmological parameters.
\end{itemize}
We are currently performing a joint analysis of the galaxy power spectra
and bispectra on the Millennium Simulation. The results will be reported
elsewhere. 

There are limitations in our present study, however. First, a relatively small
volume of the Millennium Simulation does not allow us to make a
precision test of the 3rd-order PT. Also, this limitation does not
allow us to study constraints on more than one cosmological
parameter. We have picked $D$ as the representative example because
this parameter seems the most interesting one in light of the future
surveys whose primary goal is to constrain the properties of dark
energy. In the future we must use larger simulations to show
convincingly that the bias in cosmological parameters is much lower than
1\% level. 
Second, we have found that, 
due to the limited statistics of a
small volume and the smaller $k_{max}$ due
to stronger non-linearities, the
bias parameters are not determined very well from the galaxy power
spectra alone at $z\le 3$. This issue should disappear by including the
bispectrum in the joint analysis.
Last and foremost, our study has been
restricted to the real space power spectra: we have not addressed the
non-linearities in redshift space distortion. This is a subject of the
future study.


\acknowledgments
We would like to thank Volker Springel for providing us with the matter
power spectrum data from the Millennium Simulation shown in
\S~\ref{sec:DM}, and Gerard Lemson for his help on the
Millennium database.
We would like to thank Paul Shapiro and Ilian Iliev for their
contribution during the initial stage of this project.
This material is based in part upon work supported by the Texas Advanced
Research Program under Grant No. 003658-0005-2006. 
E.K. acknowledges support from an Alfred P. Sloan Fellowship.
The Millennium Simulation databases used in this paper and the web application
providing online access to them were constructed as part of the activities
of the German Astrophysical Virtual Observatory.

\appendix
\section{Error on power spectrum}\label{sec:appA}
Besides the normalization,
an estimator for the power spectrum may be written as
\begin{equation}
P_{obs}(k)
=
\frac{1}{N_k} \sum_{i=1}^{N_k}
|\delta(\mathbf{k}_i)|^2
\Biggl|_{|\mathbf{k}_i-k|\le\Delta k}
\end{equation}
where $\delta(\mathbf{k}_i)$ is a Fourier transform of the density field
in position space, $\Delta k$ is the fundamental wavenumber of either
survey volume or simulation box, and $N_k$ is the number of
independent $k$-modes available per bin.
This estimator is unbiased because
\begin{equation}
\langle P_{obs}(k) \rangle
=
\frac{1}{N_k} \sum_{i=1}^{N_k}
\langle 
|\delta(\mathbf{k}_i)|^2
\rangle 
\Biggl|_{|\mathbf{k}_i-k|\le\Delta k}
=
\langle 
|\delta(k)|^2
\rangle
=P(k),
\end{equation}
where $P(k)$ is the underlying power spectrum.
The variance of this estimator is given by 
\begin{equation}
\left\langle
\left( \frac{P_{obs}(k)-P(k)}{P(k)} \right)^2
\right\rangle
=1-2 \frac{\langle P_{obs} \rangle}{P(k)}
+\frac{1}{N_k^2P(k)^2}\sum_{i=1}^{N_k}\sum_{j=1}^{N_k}
\langle
\delta^*(\mathbf{k}_i)\delta(\mathbf{k}_i)
\delta^*(\mathbf{k}_j)\delta(\mathbf{k}_j)
\rangle.
\end{equation}
Assuming that the density field is a Gaussian random variable 
with its variance given by $P(k)$, i.e., 
\begin{equation}
\langle\delta_i^*\delta_j\rangle =P(k)\delta_{ij},
\end{equation}
we use  Wick's theorem for evaluating the last double summation:
\begin{eqnarray}
\label{eq:wick}
\sum_{i=1}^{N_k}\sum_{j=1}^{N_k}
\langle
\delta^*_i\delta_i
\delta^*_j\delta_j
\rangle
&=&
\sum_{i=1}^{N_k}\sum_{j=1}^{N_k}
\left[
\langle
\delta^*_i\delta_i
\rangle
\langle
\delta^*_j\delta_j
\rangle
+
\langle
\delta^*_i\delta_j
\rangle
\langle
\delta^*_j\delta_i
\rangle
+
\langle
\delta^*_i\delta^*_j
\rangle
\langle
\delta_i\delta_j
\rangle
\right]\nonumber\\
&=&
N_k^2 [P(k)]^2 + N_k [P(k)]^2. 
\end{eqnarray}
Therefore, the variance is given by
\begin{equation}
\left\langle
\left[P_{obs}(k)-P(k) \right]^2
\right\rangle
= \frac{[P(k)]^2}{N_k},
\end{equation}
and the standard deviation is given by
\begin{equation}\label{eq:errpk}
\sigma_{P(k)}
\equiv
\left\langle
\left[P_{obs}(k)-P(k) \right]^2
\right\rangle^{1/2}
=
\sqrt{\frac{1}{N_k}}P(k).
\end{equation}
Note that this formula is valid only when $\delta$ is a Gaussian random
field. When $\delta$ is non-Gaussian due to, e.g., non-linear evolution,
primordial non-Gaussianity, non-linear bias, etc., we must add the
connected four-point function to Eq.~(\ref{eq:wick}).
See also \citet{takahashi/etal:prep} for the study of finite box size effects on the four-point function.

How do we calculate $N_k$? 
As the Fourier transformation of a real-valued field has symmetry given by
$\delta^*(\mathbf{k})=\delta(-\mathbf{k})$, the number of independent 
$k$-modes is exactly a half of the number of modes available in a
spherical shell at a given $k$.
We find 
\begin{equation}
N_k=\frac{1}{2} \frac{4\pi k^2\delta k}{(\delta k)^3}
=2\pi\left(\frac{k}{\delta k}\right)^2,
\end{equation}
where $\delta k$ is the fundamental wavenumber given by $\delta k =
2\pi/L$, where $L$ is the survey size or simulation box size.

In the literature one may often find a different formula such as
\begin{equation}\label{eq:errpk_litereature}
\sigma_{P(k)}^{literature}
=
\sqrt{\frac{2}{N^{literature}_k}}P(k).
\end{equation}
Here, there is an extra factor of $\sqrt{2}$, as 
$N^{literature}_k$ is the  number of modes available in a
spherical shell at a given $k$, {\it without symmetry,
$\delta^*(\mathbf{k})=\delta(-\mathbf{k})$, taken into account},
i.e., $N^{literature}_k=2N_k$. Both formulas give the same results,
provided that we understand what we mean by $N_k$ in these formulas.

\begin{figure}
\centering
\rotatebox{90}{
	\includegraphics[width=10cm]{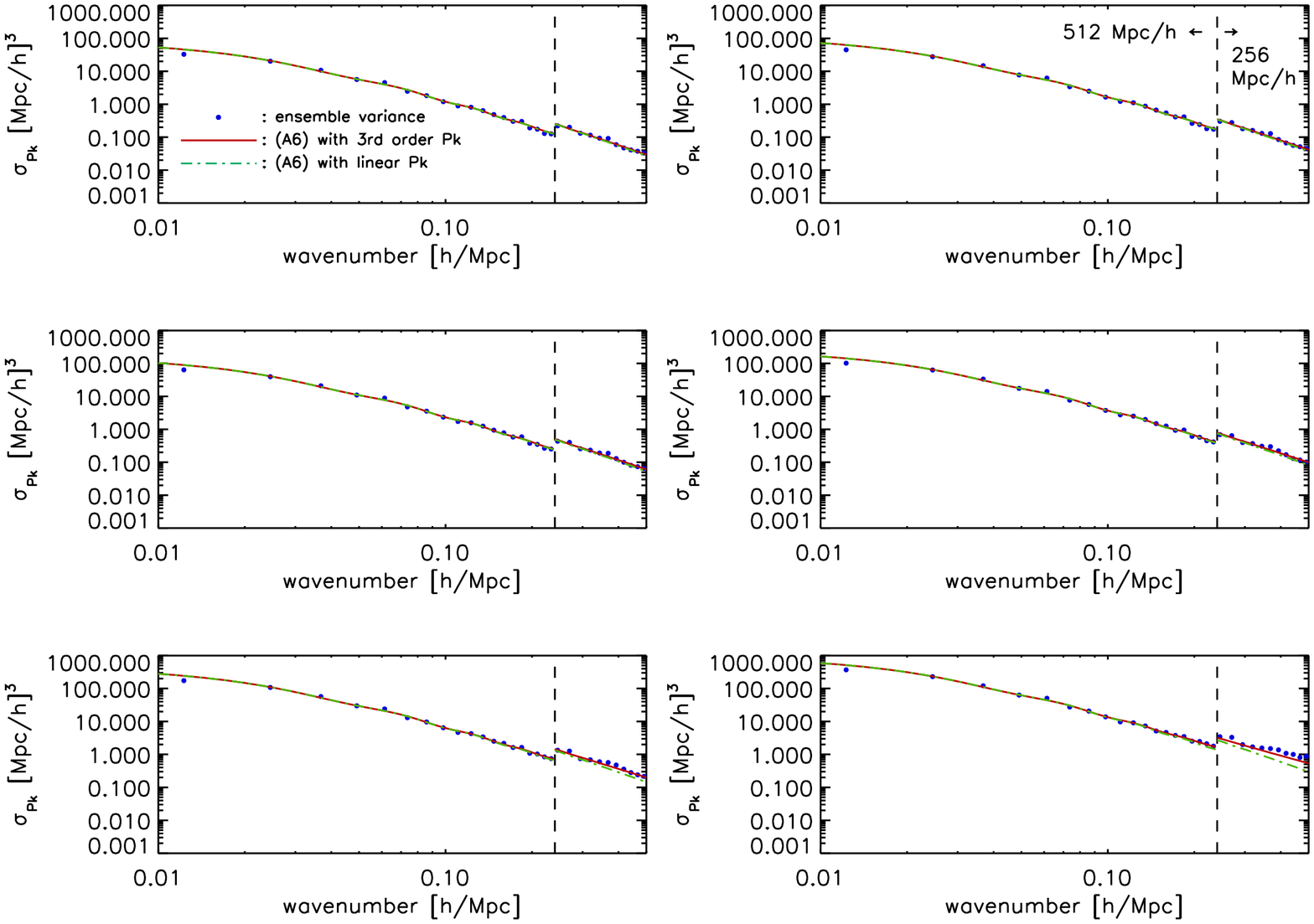}
}
\caption{
Standard deviation of the matter power spectrum: 
analytical versus simulations.
The symbols show the standard deviations directly measured from
120 independent $N$-body simulations whose box sizes are
 $L=512~\mathrm{Mpc}/h$ (60 realizations for $k<0.24 h/\mathrm{Mpc}$)
and 
 $L=256~\mathrm{Mpc}/h$ (60 realizations for $0.24<k<0.5 h/\mathrm{Mpc}$)
. Each simulation contains $256^3$ particles.
The solid and dot-dashed lines show the analytical 
formula (Eq.~(\ref{eq:errpk})) with the 3rd-order PT non-linear $P(k)$
 and the linear $P(k)$, respectively.
 Note that the graph is discontinuous at $k=0.24 h/\mathrm{Mpc}$
because the number of $k$ modes, $N_k$, for a given wavenumber $k$ 
is different for different box sizes.
}%
\label{figA1}	
\end{figure}

\begin{figure}
\centering
\rotatebox{90}{
	\includegraphics[width=10cm]{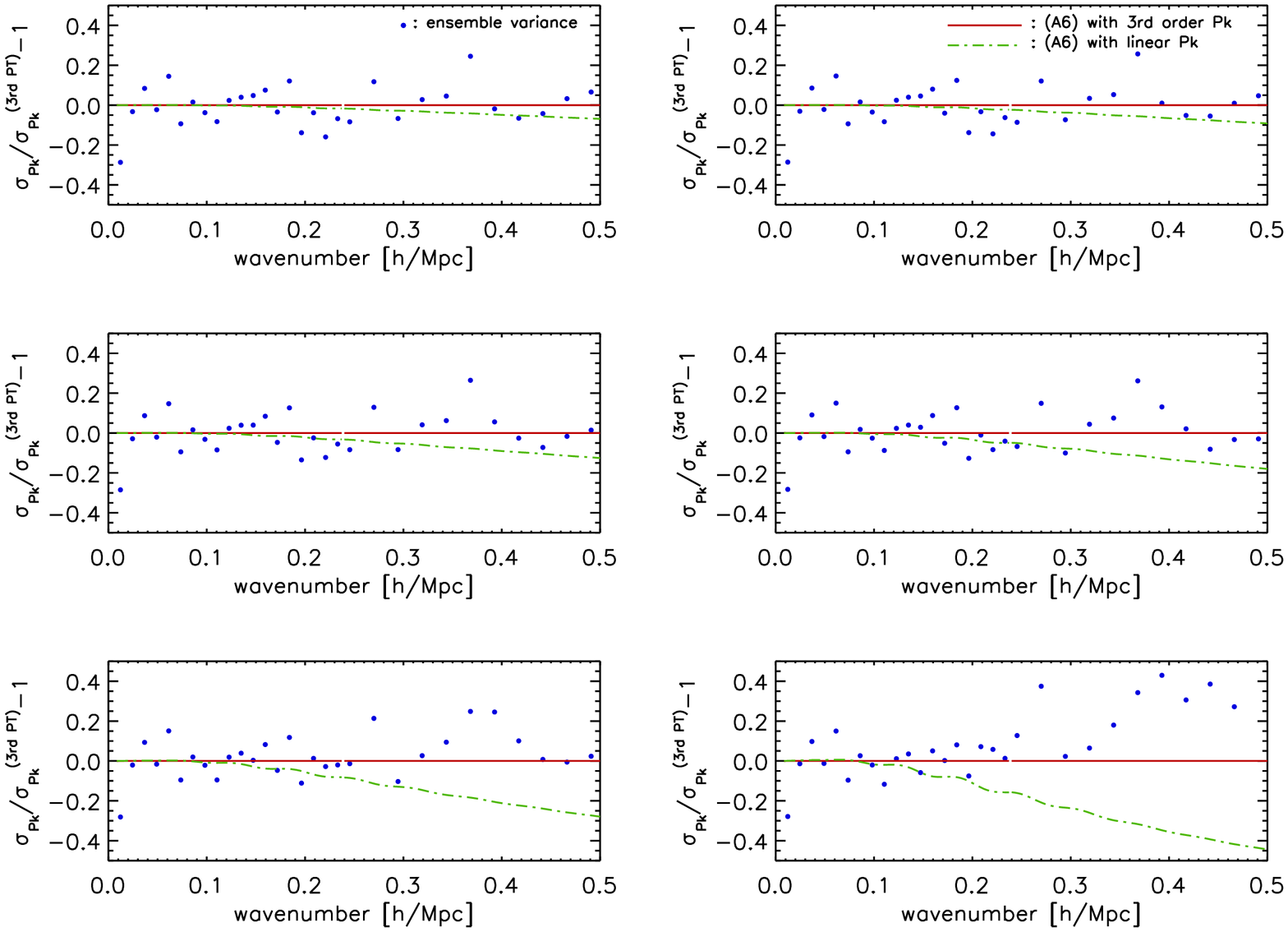}
}
\caption{
Residuals. We divide both analytical estimation and simulation results 
by the analytical formula (Eq.~(\ref{eq:errpk})) with 
the 3rd-order PT nonlinear $P(k)$.
}%
\label{figA2}
\end{figure}

We have tested the formula Eq.~(\ref{eq:errpk}) by comparing it to
the standard deviation estimated the ensemble of dark matter simulations
used in Paper I. (See Paper I for details of the simulations.)
Figure \ref{figA1} and \ref{figA2}
show the result of this comparison. 
The formula Eq.~(\ref{eq:errpk}) and the simulation data agree  well.


\section{Analytical marginalization of the likelihood function over
 $\tilde{b}_1^2$ and $P_0$}\label{sec:appB} 
In this appendix we derive the analytical formulas for the
likelihood function marginalized over $\tilde{b}_1^2$ and $P_0$.

The likelihood function, Eq.~(\ref{eq:likelihood}), is given by
\begin{equation}
\mathcal{L}(\tilde{b}_1,\tilde{b}_2,P_0,\theta_n)
=
\left(\prod_{i}\frac{1}{\sqrt{2\pi\sigma_{Pi}^2}}\right)
\exp
\left[
-
\sum_{i}
\frac
{\left(P_{obs,i}
-
\tilde{b}_1^2(P_{\delta\delta,i}+\tilde{b}_2P_{b2,i}
+\tilde{b}_2^2P_{b22,i})-P_0\right)^2}
{2\sigma_{Pi}^2}
\right],
\end{equation}
where $\theta_n$ are the cosmological parameters that do not depend on 
any of the bias parameters.
The subscript $i$ denotes bins, $k_i$.

Integrating the likelihood function over $P_0$, we find
\begin{eqnarray}\label{eq:marN}
\mathcal{L}(\tilde{b}_1,\tilde{b}_2,\theta_n)
&
=&
\int_{-\infty}^{\infty} d P_0 \mathcal{L}(\tilde{b}_1,\tilde{b}_2,P_0,\theta_n)
\nonumber
\\
&
=&
\mathcal{N}\sqrt{\frac{2 \pi}{\sum_i w_i}}
\exp\left[
-\frac{1}{2}
\frac{
\sum_{i>j} w_iw_j(a_j-a_i)^2
}
{\sum_i w_i}
\right],
\end{eqnarray}
where we have defined new variables
\begin{eqnarray}
\mathcal{N}
&\equiv&\prod_{i}
\frac{1}{\sqrt{2\pi\sigma_{Pi}^2}}\\
\label{eq:ai}a_i&\equiv&
P_{obs,i}
-
\tilde{b}_1^2(P_{\delta\delta,i}+\tilde{b}_2P_{b2,i}+\tilde{b}_2^2P_{b22,i})\\
w_i&\equiv&
\frac{1}{\sigma_{Pi}^2}.
\end{eqnarray}

We then integrate this function over $\tilde{b}_1^2$.
Introducing new variables given by
\begin{eqnarray}
\bar{\mathcal{N}}
&\equiv&\mathcal{N}\sqrt{\frac{2 \pi}{\sum_i w_i}},\\
P_{th,i}&\equiv&P_{\delta\delta,i}+\tilde{b}_2P_{b2,i}+\tilde{b}_2^2P_{b22,i},
\end{eqnarray}
and $a_i=P_{obs,i}-\tilde{b}_1^2P_{th,i}$, we rewrite Eq.~(\ref{eq:marN}) as
\begin{eqnarray}
\mathcal{L}(\tilde{b}_1,\tilde{b}_2,\theta_n)
&=&
\bar{\mathcal{N}}
\exp
\left[
-\frac{1}{2}
\frac{
\sum_{i>j} w_iw_j \left\{(P_{th,i}-P_{th,j})\tilde{b}_1^2-(P_{obs,i}-P_{obs,j})\right\}^2
}
{\sum_i w_i}
\right]
\nonumber\\
&=&
\bar{\mathcal{N}}
\exp
\left[
-\frac{1}{2}\left(A \tilde{b}_1^4-2B \tilde{b}_1^2 +C\right)
\right],
\end{eqnarray}
where
\begin{eqnarray}
A&\equiv&\frac{\sum_{i>j}w_iw_j(P_{th,i}-P_{th,j})^2}{\sum_iw_i}\\
B&\equiv&\frac{\sum_{i>j}w_iw_j(P_{th,i}-P_{th,j})(P_{obs,i}-P_{obs,j})}{\sum_iw_i}\\
C&\equiv&\frac{\sum_{i>j}w_iw_j(P_{obs,i}-P_{obs,j})^2}{\sum_iw_i}.
\end{eqnarray}
Assuming a flat prior on $\tilde{b}_1^2$, we integrate 
the likelihood function to find the desired result:
\begin{eqnarray}\label{eq:marNb1}
\mathcal{L}(\tilde{b}_2,\theta_n)
&
=&
\bar{\mathcal{N}}
\int_0^{\infty}\exp
\left[
-\frac{1}{2}\left(A \tilde{b}_1^4-2B \tilde{b}_1^2 +C\right)
\right]
d(\tilde{b}_1^2)
\nonumber\\
&
=&
\bar{\mathcal{N}}
\exp\left[
\frac{B^2-AC}{2A}
\right]
\sqrt{\frac{\pi}{2A}}\left\{1+\rm{erf}\left(\frac{B}{\sqrt{2A}}\right)\right\}. 
\end{eqnarray}
Note that the convergence of the likelihood function is ensured 
by Cauchy's inequality, $B^2-AC<0$.

\section{distribution of errors on the distance scale}
\label{sec:appC} 
We find that the error on $D$ extracted from the halo power spectrum 
of Millennium Simulation is about $2.17\%$ for $k_{max}=1.5~h/\mathrm{Mpc}$
at $z=6$. (See Figure \ref{fig16}.)
On the other hand, the error on $D$ calculated from the Fisher information
matrix is $1.57\%$
Are they consistent?

In order to test whether it is possible to get the error on $D$ far from
the value derived from the Fisher matrix, we generate $1000$ realizations
of mock power spectra with the best-fitting bias parameters for halo
with $k_{max}=1.5~h/\mathrm{Mpc}$ at $z=6$.
Then, we calculate the best-fitting value of $D$ as well as the 1-$\sigma$ 
($68.27\%$ CL) range from the one-dimensional marginalized likelihood
function of $D$ for each realization. 

\begin{figure}
\centering
\rotatebox{90}{
	\includegraphics[width=10cm]{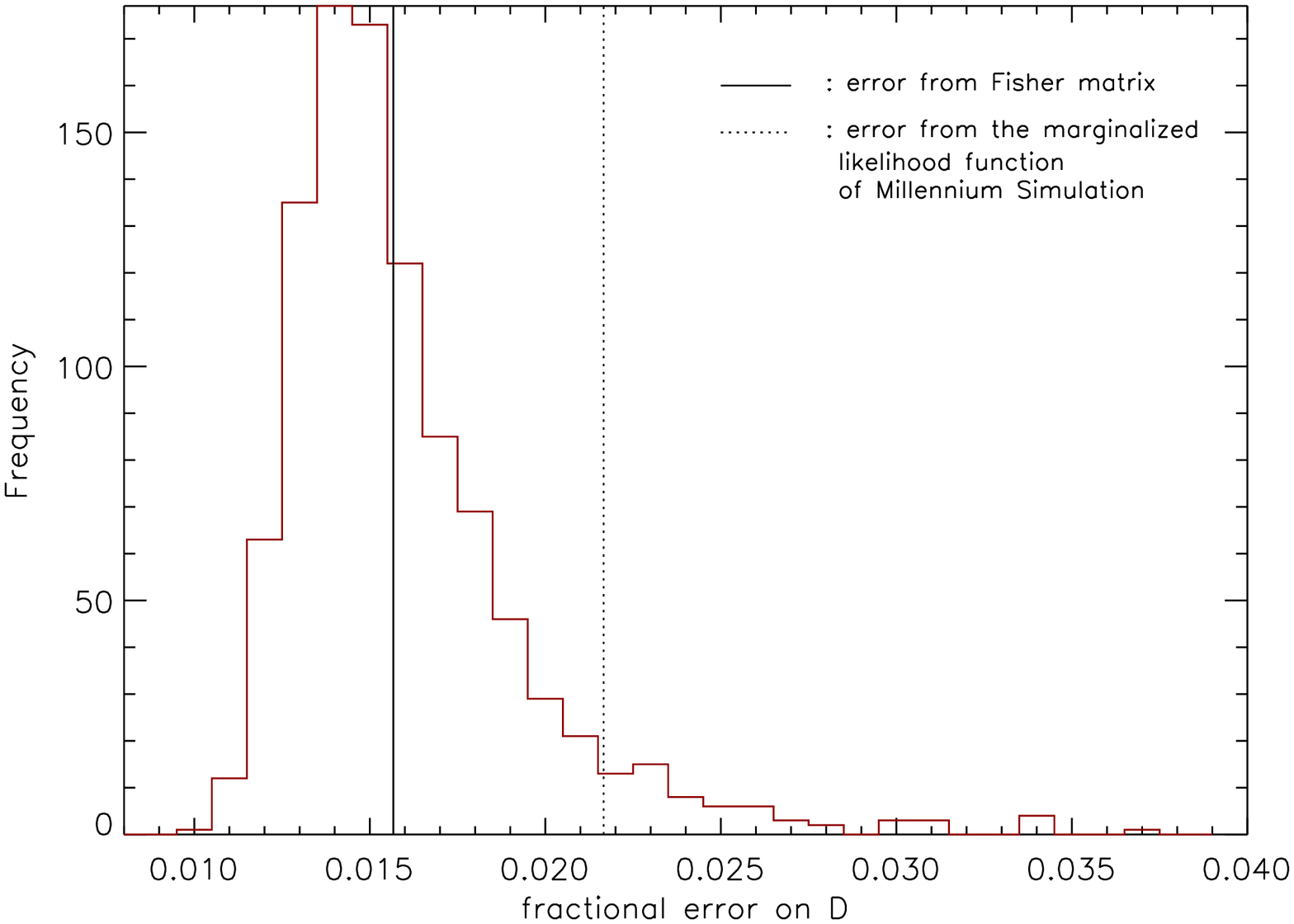}
}
\caption{
Histogram for the 1-$\sigma$ errors on $D$ calculated from 
$1000$ Monte Carlo realizations generated
with the best-fitting bias parameters of halo power spectrum of
Millennium Simulation with $k_{max}=1.5~h/\mathrm{Mpc}$ at $z=6$.
The error derived from the Fisher matrix is close to the mean, 
while the error from the marginalized one-dimensional likelihood function
of Millennium Simulation is on the tail of the distribution.
The probability of having an error on $D$ greater than that from the 
Millennium Simulation is about $6 \%$.
}%
\label{figC1}
\end{figure}

We find that the mean 1-$\sigma$ error on $D$ calculated 
from these realizations is $1.66\%$, and their standard deviation is $0.43\%$.
Figure \ref{figC1} shows the distribution of the fractional 
1-$\sigma$ error on $D$ compared with $D_{true}$. 
While the error derived from the Fisher matrix is close to the mean,
the error calculated from the Millennium Simulation is on the tail of the
distribution. The probability of having an error on $D$ greater 
than that from the Millennium Simulation is about $6 \%$,
which is acceptable.

\end{document}